\RequirePackage{fix-cm}

\documentclass[twocolumn,epjc3]{svjour3}  

\RequirePackage{graphicx}
\smartqed  
                             
\usepackage{epsfig,dcolumn}
\usepackage{graphicx}
\usepackage{color}                 
\usepackage{amsmath}
\usepackage{amsfonts}
\usepackage{amssymb}
\usepackage{graphicx}
\usepackage{widetext}
\usepackage{type1cm}
\usepackage{eso-pic}

\newcommand{\be}{\begin{equation}}
\newcommand{\ee}{\end{equation}}

\newcommand{\im}{\mathrm{Im}\,}
\newcommand{\re}{\mathrm{Re}\,}

\usepackage{bm} 

\journalname{Eur. Phys. J. C}

\begin{document}

\title{Strange resonance poles from $K\pi$ scattering below 1.8 GeV}

\author{J.R.~Pelaez\thanksref{addr1}
        \and
        A.Rodas\thanksref{addr1}
				\and
		    J.Ruiz de Elvira\thanksref{addr2}		
}				
				
				\institute{Departamento de F\'isica Te\'orica II, Universidad Complutense de Madrid, 28040 Madrid, Spain \label{addr1}
           \and
Helmholtz--Institut f\"ur Strahlen- und Kernphysik (Theorie) and
   Bethe Center for Theoretical Physics, Universit\"at Bonn, D--53115 Bonn, Germany \label{addr2}\\
Albert Einstein Center for Fundamental Physics, Institute for Theoretical Physics,
University of Bern, Sidlerstrasse 5, 3012 Bern, Switzerland
}

\date{Received: \today / Accepted: \today}

\maketitle

\begin{abstract}
In this work we present a determination of the mass, width, and coupling of
the resonances that appear in kaon-pion scattering below 1.8 GeV.
These are: the much debated scalar $\kappa$-meson, nowadays known as $K_0^*(800)$,
the scalar $K_0^*(1430)$, the $K^*(892)$ and $K_1^*(1410)$ vectors,  the spin-two
$K_2^*(1430)$ as well as the spin-three $K^*_3(1780)$. The parameters will be determined 
from the pole associated to each resonance by means of an analytic continuation
of the $K\pi$ scattering amplitudes obtained in a recent and precise data analysis
constrained with dispersion relations, which were not well satisfied in previous analyses. 
This analytic continuation will be performed 
by means of Pad\'e approximants, thus avoiding a particular model for the
pole parameterization. We also pay particular attention to the evaluation of uncertainties.
\end{abstract}
\maketitle

\section{Introduction}

A reliable determination of strange resonances 
is by itself relevant for hadron spectroscopy and their own
 classification in multiplets, 
 as well as for our understanding of intermediate energy QCD and the low-energy regime through Chiral 
Perturbation Theory. In addition kaon-pion scattering and
 the resonances that appear in it are also of interest
because most hadronic processes with net strangeness end up with at least a $K\pi$ pair
that contributes decisively to shape the whole amplitude through final 
state interactions. This is, for instance, the case of heavy B or D meson decays into kaons and pions. Actually, the parameterization of these amplitudes and their final states interaction is very frequently done in terms of simple resonance exchange models. Conversely, although many of the strange resonances were observed in $K\pi$ scattering long ago \cite{Aston:1986rm}, most of them were later confirmed in studies of heavier meson decays, which were also used to determine their parameters.

However, very often the analyses of these resonances have been made 
in terms of crude models,
which make use of specific parameterizations like isobars, 
Breit--Wigner forms or modifications, which very often assume the existence of some simple background.
As a result, resonance 
parameters suffer a large model dependence or may
even be process dependent. Thus, the statistical uncertainties 
in the resonance parameters should be accompanied by 
systematic errors that are usually ignored. This can easily be checked by looking at the Review of Particle Physics (RPP) compilation \cite{PDG}, where very frequently 
for these resonances it is only possible to provide an ``estimate'' of their mass or width, 
together with some 
educated guess for the uncertainty, since
the central values reported by different experiments on the same resonance are inconsistent
among themselves. Part of these discrepancies may  definitely be due to systematic effects on data, 
but 
to a large extent they are due to the use of models in their analysis to extract resonance parameters. 
In some cases, as for the $K_0^*(800)$, even the very existence of the resonance is called for confirmation. 

The most rigorous way of identifying the parameters of a resonance 
is from the position  $s_R$ of its associated pole  
in the complex energy-squared plane, which is related to 
the resonance mass $M_R$ and width $\Gamma_R$ by $\sqrt{s_R}\equiv M-i\Gamma/2$.
The reason is that poles are process independent, whereas determining
resonance parameters from peaks or bumps on the data depends on backgrounds as well as on
 the presence of thresholds or other resonance contributions specific to each process.
 
But even when using the pole definition there is an additional problem;
the data can be equally well described in a given region by different functional forms
whose analytic continuation is different. For instance, in a given energy interval, 
data could be fitted with a polynomial 
of sufficiently high degree, and such a parameterization never has a pole nor cuts. If the resonance is narrow and isolated
one can use physically motivated functions like a Breit-Wigner formula or variations.
However, as soon as resonances are wide and their associated poles require 
an analytic continuation deep in the complex plane or if there are coupled channels 
with thresholds nearby  or overlapping resonances,
 it is better to avoid models for the analytic continuation to the pole.

The most rigorous way to determine poles in the complex plane
is to perform an analytic continuation of the amplitude
by means of partial-wave dispersion relations~\cite{Ananthanarayan:2000ht,Buettiker:2003pp,GarciaMartin:2011cn,Hoferichter:2015hva}. 
A paradigmatic example has been the recent determination of
the long debated $\sigma/f_0(500)$ pole by means of Roy \cite{Roy:1971tc} 
and GKPY equations \cite{GarciaMartin:2011jx}, which triggered a 
radical revision of its parameters in the RPP (see for a detailed account
of this progress \cite{Pelaez:2015qba}).
However, although a similar dispersive analysis for the $K_0^*(800)$ 
in terms of Roy-Steiner equations
has been performed \cite{DescotesGenon:2006uk}, the $K_0^*(800)$ status in the RPP 
is that it still ``Needs confirmation''. 
These partial wave dispersion relations are very rigorous and take into account
the contributions from all the singularities in the complex plane and particularly
those of the left-hand cut due to thresholds in crossed channels.
The price to pay is that they  are 
complicated sets of coupled integral equations whose convergence region
in the complex plane only covers the lowest resonances. Moreover, they use as input
waves beyond $J=1$ as well as in
the intermediate energy region, which typically includes the inelastic region.
Therefore, in practice, the amplitudes obtained in these studies only satisfy precisely 
these partial-wave dispersive constraints up to energies slightly beyond the elastic regime, at best.
In our case this makes 
them valid to study the $K_0^*(800)$ and  $K_0^*(892)$, but unsuitable 
to determine the parameters of all the other resonances
appearing in $K\pi$ scattering below 1.8 GeV. 
Hence, the use of dispersion relations to make rigorous analytic continuations
of partial waves to the complex plane is therefore rather limited 
for resonances well above 1 GeV.

For the above reasons there is a growing interest
in other methods based on analyticity properties 
to extract resonance pole parameters
from data in a given energy domain.
They are based on several approaches: conformal expansions
to exploit the maximum analyticity domain of the amplitude 
\cite{Cherry:2000ut,Yndurain:2007qm,Caprini:2008fc},
Laurent \cite{Guo:2015daa} or Laurent-Pietarinen \cite{Svarc:2013laa} expansions,
or Pad\'e approximants \cite{Masjuan:2013jha,Masjuan:2014psa,Caprini:2016uxy}.
They all determine the pole position without assuming a particular 
model for the relation between the mass, width and residue.
In this sense they are model independent analytic continuations to the complex plane.

Of course, these analytic methods require as input
some data description. But it is not enough that it may be a precise description:
 it should also be consistent with some basic 
principles, which usually is not the case.
Actually, it has been recently shown \cite{Pelaez:2016tgi} that $K\pi$ scattering data \cite{Estabrooks:1977xe,Aston:1987ir}, which are the source for several determinations of strange resonances, 
do not satisfy well Forward Dispersion Relations up to 1.8 GeV.
This means that in the process of extracting data by using models,
they have become in conflict with causality.
Nevertheless, in \cite{Pelaez:2016tgi}
the data were refitted 
constrained to satisfy those Forward Dispersion Relations and 
a careful systematic and statistical error analysis was provided.  
The constrained fits suffer some visible 
changes compared to unconstrained fits
and is therefore of interest to check the resonance parameters resulting from this constrained analysis.
In this work we will make use of the Pad\'e approximants method 
in order to extract the parameters of all resonances appearing in those waves.

The plan of this article is as follows. In the next section we will briefly review the status of 
data for $K\pi$ scattering 
and their phenomenological description. Then, in Sec.~\ref{sec:intropades} 
the Pad\'e approximant method will be introduced. In Sec.\ref{sec:results} we present
our numerical results in separated subsections dedicated to 
scalar, vector and tensor resonances. Finally, in Sec.~\ref{sec:conclusions} we provide our conclusions.

\section{$K\pi$ scattering}
\label{sec:Kpi}

Data on $K\pi$ scattering were measured indirectly from $KN\rightarrow K\pi N$ reactions
during the 70s and the 80s. The most widely used are those of 
Estabrooks et al. \cite{Estabrooks:1977xe} and Aston et al. \cite{Aston:1987ir}, 
 which provide amplitude phases and modulus up to roughly 1.8 GeV.
Note they are all extracted within an isospin limit formalism, so that charged and neutral 
mesons are assumed to have the same mass. Here we will use $m_\pi=139.57\,$MeV and $m_K=496\,$MeV.

Apart from the simple phenomenological parameterizations 
of the original experimental articles \cite{Estabrooks:1977xe,Aston:1986rm,Aston:1987ir}, the data set, or parts of it,
has been described with a wide variety of approaches, 
also used to identify strange resonances below 1.8 GeV. 
For instance, already in the 80s the $S$-wave was described up to almost 1.3 GeV with a unitarized model of mesons coupled to quark-antiquark confined channels \cite{vanBeveren:1986ea}. 
In the 90s, the $S$ and $P$ waves
were described
with unitarized Chiral Perturbation Theory, using 
the Inverse Amplitude Method first in the elastic regime
\cite{Dobado:1992ha} and then with coupled channels up to 1.2 GeV \cite{GomezNicola:2001as}.
An alternative unitarization method for ChPT amplitudes
described $S$-wave data up to 
1.43 GeV \cite{Zheng:2003rw}. In addition, data has also been described  
with: i) the chiral unitary approach
 to  next to leading order \cite{Oller:1997ng} for the $S$ and $P$-waves, 
ii) the N/D unitarization 
approach with coupled channels for the $S$-wave up to 1.4 GeV,
iii) unitarized chiral Lagrangians that 
include some resonances explicitly while others are generated dynamically 
for the $S$ wave \cite{Black:1998zc,Jamin:2000wn,Wolkanowski:2015jtc,Ledwig:2014cla}, 
iv) conformal parameterizations \cite{Cherry:2000ut} for the $S$-wave, v) the
explicit consideration of resonances with ad-hoc pole parameterizations and 
very simple chiral symmetry requirements for the $S$-wave 
\cite{Ishida:1997wn,Bugg:2003kj}, or vi) 
unitarized models with resonances \cite{Magalhaes:2014ova} for the $P$-wave.
Note that these models do not deal with $D$ or $F$ waves.

Not all those models are equally rigorous, but 
in all them partial-wave unitarity plays a central role. 
The most constrained by fundamental
principles are those including chiral symmetry constraints and based 
on dispersion relations, although usually they have
some approximation for the so-called left-hand and circular cuts, 
which are branch cuts due to 
thresholds in crossed channels or to the angular integration of Legendre polynomials. The most rigorous treatment is the Roy-Steiner equation analysis of \cite{Buettiker:2003pp,DescotesGenon:2006uk} for the $S$ and $P$-waves, where left and circular cuts are treated systematically,
although it only extends to energies below $\sqrt s\simeq0.97$ GeV and the amplitudes above that energy or higher angular momentum are considered input.

It is very important to remark that all the approaches above  make use of the existing scattering data from 
\cite{Estabrooks:1977xe} and \cite{Aston:1987ir}.
However, for the extraction of those $K\pi$ scattering data 
 from $KN\rightarrow K\pi N$, 
several approximations and assumptions were needed.
For instance, it was assumed that the full process 
is dominated by one pion exchange (OPE-model),
frequently neglecting final
rescattering with the nucleon or the exchange of other resonances. 
In addition the OPE was approximated by an on-shell extrapolation.
These are sources of systematic uncertainty, not directly provided in the experimental papers,
which  explain in part why different experiments do not always agree within their quoted
uncertainties, which are of statistical nature. 
As a matter of fact, it has been recently shown \cite{Pelaez:2016tgi}
that simple fits to those data do not satisfy well Forward Dispersion Relations (FDR) up to 1.8 GeV, 
even when including estimates of the 
systematic uncertainty (typically estimated as the difference between conflicting data points).
Note that, since the Roy-Steiner formalism
is in practice limited to energies below $\sqrt s\simeq$ 1 GeV,
above that energy it is only possible to test two independent FDRs.

Nevertheless, the existing data was also refitted in \cite{Pelaez:2016tgi},
but constrained to satisfy FDRs. The resulting Constrained Fit to Data (CFD)
provides a precise description of data, which is consistent within uncertainties
with two FDRs, although only up to 1.6 GeV.
The CFD is a rather simple set of parameterizations of the $S$, $P$, $D$ and $F$ partial-wave
phase shifts and inelasticities in the isospin limit, 
for both possible isospins $I=1/2$ and $3/2$, as well as a Regge description above 1.7 GeV.
These parameterizations are
given as piecewise functions. Each piece is valid in a given 
energy interval of real energies and is matched continuously to the next piece, 
typically at 
different energy thresholds. No model dependent assumptions are thus made.

However, these parameterizations should not 
be used directly to extract resonance parameters.
The functional form of each piece of those parameterizations has been chosen 
to be simple and flexible enough to describe the amplitude
in a certain interval of real energies.
Of course, each 
piece of function by itself may be continued to the complex plane in 
a certain domain that depends on the analytic structure of that piece.
However, 
that analytic extension is not necessarily a good approximation to the
continuation of the whole amplitude to the complex plane, which has a definite 
analytic structure in terms of cuts associated to physical thresholds.

This is rather general, not just an issue with the CFD, since one could always fit peaks and 
dips in a finite energy interval with a polynomial, 
whose analytic continuation 
would never provide a pole in the complex plane.
The same happens with a Breit-Wigner formula, which can always be fitted to 
a peak in an interval, with some choice of smooth background if needed. This always produces a pole, 
but it only has some physical meaning if the pole is close to the real axis and well isolated from other singularities. Note that this parameterization or any of its modifications 
(with kinetic factors or Blatt-Weiskopf barrier factors) 
also imposes a particular relation between the pole position and residue.

Thus, in order to extract pole parameters from the Constrained Fit to Data in  \cite{Pelaez:2016tgi} we will make use of the Pad\'e method, which extracts the pole in a given interval once the analytic structure in a domain that contains the pole of the resonance is fixed, without imposing a particular relation between the position and residue of that resonance.

\section{Pole determination using Pad\'e approximants}
\label{sec:intropades}

The $P^N_M(s,s_0)=Q_N(s,s_0)/R_M(s,s_0)$  Pad\'e approximant 
of a function $F(s)$ is a rational function that satisfies
\begin{equation}
P^N_M(s,s_0)=F(s)+O((s-s_0)^{M+N+1}),
\end{equation}
with  $Q_N(s,s_0)$ and $R_M(s,s_0)$ polynomials in $s$ of order $N$ and $M$, respectively.
These approximants 
can be calculated easily from the derivatives of the data 
fit with respect to the energy squared $s$. 

Thanks to the de Montessus de Ballore theorem these Pad\'e approximants 
can be used to unfold the next continuous Riemann sheet of a scattering amplitude
in order to search for resonance poles 
\cite{Masjuan:2013jha,Masjuan:2014psa,Caprini:2016uxy}. 
The relevant observation is that when 
they yield a pole they do not assume a model for the relation between
its position and residue. Hence, in this sense the pole is model independent, although
there is some residual dependence on the choice of parameterization for the data,  
from which the derivatives are obtained \cite{Caprini:2016uxy}. 
This will be taken into account into our systematic error estimation.

The choice of Pad\'e series to be used, with more or less poles,
is based on the expected analytic structure of the partial wave
in a domain that includes a segment in the real axis and the
pole of the resonance we study. 
Therefore, the series should have at least a pole 
to describe the resonance, but if in order to contain that pole the domain also
contains another singularity, like a branch point, we will need a series with 
an additional pole. 

For example, when the resonance is narrow and well isolated from other singularities,
the amplitude $F(s)$ must be analytic inside a domain around
a real $s_0$,
except for a single pole at $s=s_p$. Note that the upper half of this domain lies
on the first, or ``physical'', Riemann sheet and has no poles. 
In contrast, the lower half lies on the unphysical Riemann
sheet that is connected continuously with the first when crossing the real axis and thus it can contain poles.
In such case we can use the sequence
\begin{equation}
P^N_1(s,s_0)=\sum^{N-1}_{k=0}{a_k(s-s_0)^k+\frac{a_N(s-s_0)^N}{1-\frac{a_{N+1}}{a_N}(s-s_0)}},
\end{equation}
which converges to $F(s)$ within the domain of analyticity excluding $s_p$. The constants $a_n=\frac{1}{n!}F^{(n)}(s_0)$ are given by the $n^{th}$ derivative of the function. 
This is how an analytic continuation to the complex plane can be 
obtained just from the fit of a function $F(s)$
to the data in the physical region of the real $s$ axis. Likewise, the pole and residue are
\begin{equation}
s_p^{N}=s_0+\frac{a_N}{a_{N+1}},~Z^{N}=-\frac{(a_N)^{N+2}}{(a_{N+1})^{N+1}}.
\label{eq:poleresidue}
\end{equation}

Note that the coupling of a given resonance 
to $K \pi$ can be obtained from the residue as follows:
\begin{equation}
\vert g_{K \pi}\vert^2=\frac{16\pi(2l+1)|Z|}{(2q_{K \pi}(s_p))^{2l}},
\label{eq:coupling}
\end{equation}
where
\begin{equation}
q_{K \pi}(s)=\frac{1}{2}\sqrt{\frac{(s-(m_{K}+m_{\pi})^2)(s-(m_{K}-m_{\pi})^2)}{s}}\nonumber
\end{equation}
is the center-of-mass momentum of the $K \pi$ system and
$l$ the angular momentum of the partial wave.

However, when the pole associated to a resonance lies near a branch cut produced by unitarity,
we may need one additional pole to mimic the branch points
inside the domain.
In such cases we will use the following sequence with $M=2$: 

\begin{widetext}
\begin{eqnarray}
P^N_2(s,s_0)=\frac{\sum^N_{k=0}{(a_ka_N^2-a_ka_{N-1}a_{N+1}-a_{k-1}a_{N}a_{N+1}+a_{k-1}a_{N-1}a_{N+2}+a_{k-2}a_{N+1}^2-a_{k-2}a_Na_{N+2})(s-s_0)^k}}{a_N^2-a_{N-1}a_{N+1}+(a_{N-1}a_{N+2}-a_Na_{N+1})(s-s_0)-(a_Na_{N+2}-a_{N+1}^2)(s-s_0)^2}, \nonumber\\
\qquad
\end{eqnarray}
\end{widetext}
\noindent
which has similar converge properties. The explicit expression for the poles may be found in \cite{Caprini:2016uxy}.
In the case of the $\kappa$, when using this $M=2$ sequence of Pad\'e series, 
we will see that one of the poles will converge to the pole associated
to the resonance $s_p$, whereas the other will simulate a branch cut.

Let us now comment on the uncertainty estimates.
From the above definitions it is clear that the calculation of pole parameters
relies on the data fitting function and its derivatives at a given energy point $s_0$. 
Thus, a first source of uncertainty is inherent to the data uncertainties
and we will refer to it as ``statistical'' error. 
We will estimate this uncertainty 
by a Monte Carlo Gaussian sampling of the fit parameters within their error bars.
Note that following the $\pi\pi$-scattering analysis in~\cite{Perez:2015pea}, 
the gaussianity of the uncertainties in the CFD was also checked in \cite{Pelaez:2016tgi}, 
hence ensuring that the standard approach for error propagation can be used.

As a second source of uncertainty, we will have a ``theoretical'' uncertainty 
due to the numerical procedure and the 
fact that the sequence of Pad\'e approximants with fixed order $M$, will be 
truncated at a given value $N$.
de Montessus de Ballore theorem tells us that,
if the amplitude in that domain and the Pad\'e series used 
for the approximation have the same number of poles, 
the differences between the $\sqrt{s^{N}_p}$  should
become smaller and the pole position should converge to
\begin{equation}
\sqrt{s_p}=M-i\Gamma/2.
\end{equation}
We thus estimate
the uncertainty in this truncation by
\begin{equation}
\Delta \sqrt{s_p^N}=\left|\sqrt{s_p^N}-\sqrt{s_p^{N-1}}\right|.
\end{equation}
We will truncate the sequence at a value of $N$ such that this error
is negligible or smaller than the ``statistical'' error.
This last $\Delta \sqrt{s_p^N}$ will then be called $\Delta_{th}$.
The center of the domain, $s_0$, is chosen as the point where this theoretical uncertainty is smaller.

Finally, we will also consider different parameterizations 
fitted to the very same CFD amplitudes described in the previous paragraph.
Note that each parameterization 
is allowed to have its own $s_0$.
Although all these parameterizations will lie within the uncertainties of the CFD in the 
real axis, they yield slightly different derivatives that result in different central values for the pole. Our final result will then 
be the average of the different values obtained
with different parameterizations and we will consider an additional systematic
uncertainty, defined as the variance of these results, due to  the model dependence when calculating the derivatives at a given point. For example, if we obtain  
values $M_i$ for the pole mass from $n$ different models, 
our final value will be the averaged mass $\bar M$ and
the systematic uncertainty will be $\Delta_{sys} M= \sqrt{\sum_i^n (M_i-\bar M)^2/(n-1)}$. 
Typically we will study other conformal parameterizations
with different conformal variables, or popular parameterizations 
like Breit-Wigners, or when these are not the most suitable choice, 
other parameterizations already used in the literature.

Our final uncertainty will be the quadratic combination
of the theoretical, statistical and systematic errors.
Similar definitions hold for the central values and systematic uncertainties for the width and coupling of the pole.

Thus, in the next sections we will show 
that we can use the sequence $P^N_1(s,s_0)$ 
to determine the poles of all strange resonances below 1.8 GeV
except for the $K_0^*(800)$. 
In these cases we truncate at $N=4$.  
For the $K_0^*(800)$ the sequence with $M=1$ 
does not converge properly to the pole position since there
is a nearby threshold which is as closer to the center of the domain
than the $\kappa$ pole itself. 
In contrast, the sequence with $M=2$ does converge rapidly to a resonance pole, 
while the other pole mimics the $K\eta$ threshold and cut.
In this case the systematic error is small enough for $N=3$. 
As a side remark, let us
note that the above Pad\'e sequence that we will use in this work
{\it  should  not be confused} with the use of a 
Pad\'e approximant to restore unitarity
on the Chiral Perturbation Theory (ChPT) expansion \cite{UChPT,Dobado:1996ps}. 
These uses of Pad\'e series
are completely unrelated. We have nevertheless found such a confusion often
and we will try to clarify this issue here. 

In the approach of this work there will be no dynamical input, only a 
parameterization of the data by means of different functions and the assumption 
that there is at least a pole in the vicinity of a certain point $s_0$. 
Using only the data description as input, 
in particular the derivatives of the amplitude at that point,
there is a series of Pad\'e approximants 
that reproduce that pole.
The only analytic structure of relevance 
is the pole and possibly some cut nearby, but the latter
 would be mimicked by further poles in the Pad\'e sequence.
Note that these Pad\'e approximants are built from a series in powers of $(s-s_0)$, that 
could be applied for any function describing data.
The inputs are only the derivatives of the amplitude at that point. 
There are no requirements from any kind of dynamics, 
particularly chiral dynamics: it is just data.
Our results in this paper will be consistent with QCD dynamics, and chiral dynamics in particular,
as long as the data are consistent with it. 

In contrast, in the case of Pad\'e series for ChPT, besides the fit to data,
there is an attempt to describe the dynamics from the ChPT Lagrangian, 
which is a low energy expansion with the QCD symmetry constraints in terms of pions, kaons and etas.
ChPT produces a series in powers of $(k/f_\pi)^2$, where $f_\pi$ is the pion decay constant and $k$ is either the meson momenta or any of their masses. 
Being organized basically as a polynomial in momentum/mass variables
the ChPT series cannot satisfy unitarity, which is a 
condition on the right-hand or physical cut. 
However, it can be shown that unitarity fixes
the imaginary part of the inverse amplitude 
on the physical or right-hand cut. Next, by using the ChPT
expansion to calculate the real part of the inverse amplitude, 
one ends up {\it formally} with a Pad\'e approximant in the 
$1/f$ expansion. 
But rigorously it is {\it not} a Pad\'e approximant in the energy or mass expansion. 
Therefore the series upon which the Pad\'e seires are built is completely different
from the one used in this work, 
and the center of the expansion is also completely different.
When used up to a given order in ChPT, 
the Pad\'e approximant ensures unitarity and, if re-expanded,
it reproduces the chiral logarithms associated to unitary of the next order in the ChPT expansion (nor the polynomial terms or the crossing logarithms of the next order \cite{Gasser:1990bv,Dobado:1996ps}).
These Pad\'e series built as resummations of the $1/f^2$ ChPT expansion are completely different from 
those used here. For further details we refer the reader to \cite{Dobado:1996ps}.
Nevertheless, the parameterizations 
we use for our central values also have a factor to account for the 
Adler zero (at leading order within ChPT) that appears below threshold in the scalar wave. 
This makes the parameterization consistent with Chiral Symmetry, 
but we could have also used here
a functional form without it, as long as it describes the data, 
since for our method we only require input around one energy point in the data region.

In summary, the approach of this work 
has absolutely nothing to do with unitarized Chiral Perturbation Theory and the Pad\'e approximants used in that case.
Quite the contrary, here we do not have any dynamical input at the Lagrangian level, and this is done on purpose, to avoid as much as possible any model dependence. We only use data as input. Of course, we use a dispersive description (again, not dynamical) of data, which has been constrained 
to satisfy forward dispersion relations, although respecting unitarity (by being parameterized only in terms of the phase-shift and inelasticity) and respecting
within uncertainties analyticity and crossing constraints. Our Pad\'e series here is just 
a consequence of the analyticity of the amplitude,
which allows for a Pad\'e expansion around a point $s_0$ that also encloses the possible resonance pole.

\section{Results}
\label{sec:results}
Let us then discuss our results for each channel.

\subsection{Scalar resonances}

In the scalar channel there are two resonances with isospin zero: the $K_0^*(800)$, which according to the RPP still ``Needs confirmation'', and the $K_0^*(1430)$. 
We start discussing the former

\subsubsection{The $K_0^*(800)$ or $\kappa$ resonance}

This resonance appears in the
low-energy region, where the scattering is still elastic. 
Note that the 
CFD parameterization describes the elastic region 
by means of a relatively simple conformal expansion 
whose explicit expression can be found in \cite{Pelaez:2016tgi}.
The advantage of such a  conformal parameterization is that once the elastic cut is separated exactly by unitarity, it provides a 
rapidly convergent expansion analytic in the whole complex plane. 
Of course, it only represents well the physical amplitude at low energies, but these good analytic properties 
already made it possible in \cite{Pelaez:2016tgi} to provide the parameters
of the pole that appears in this parameterization:
$M=680\pm15\,$MeV,  $\Gamma=668\pm15\,$MeV and $g_{K_0^*(800)K\pi}=4.99\pm0.08\,$GeV.
If we only use input from the elastic region, the Pad\'e approach should in principle
reproduce this pole at that position and therefore the present analysis for this resonance
would be of limited value.
However, the fact that we already have a precise determination of the pole 
will be useful to calibrate and understand the uncertainties
of the Pad\'e approach due to the truncation of the series and the use of different 
data parameterizations
to calculate the derivatives, or to illustrate how to choose the center of the expansion
and the most convenient Pad\'e series.
In particular, since this resonance has such a large width,
one would need to reach deep in the complex plane and 
it is likely that the Pad\'e sequence will be sensitive 
to other singularities, particularly to thresholds nearby.
Actually we will see that in this case the $M=1$ Pad\'e series, which only
has one pole, will not converge and we will need the $M=2$ series.

\begin{figure}
\centering
\centerline{ \includegraphics[width=1\linewidth]{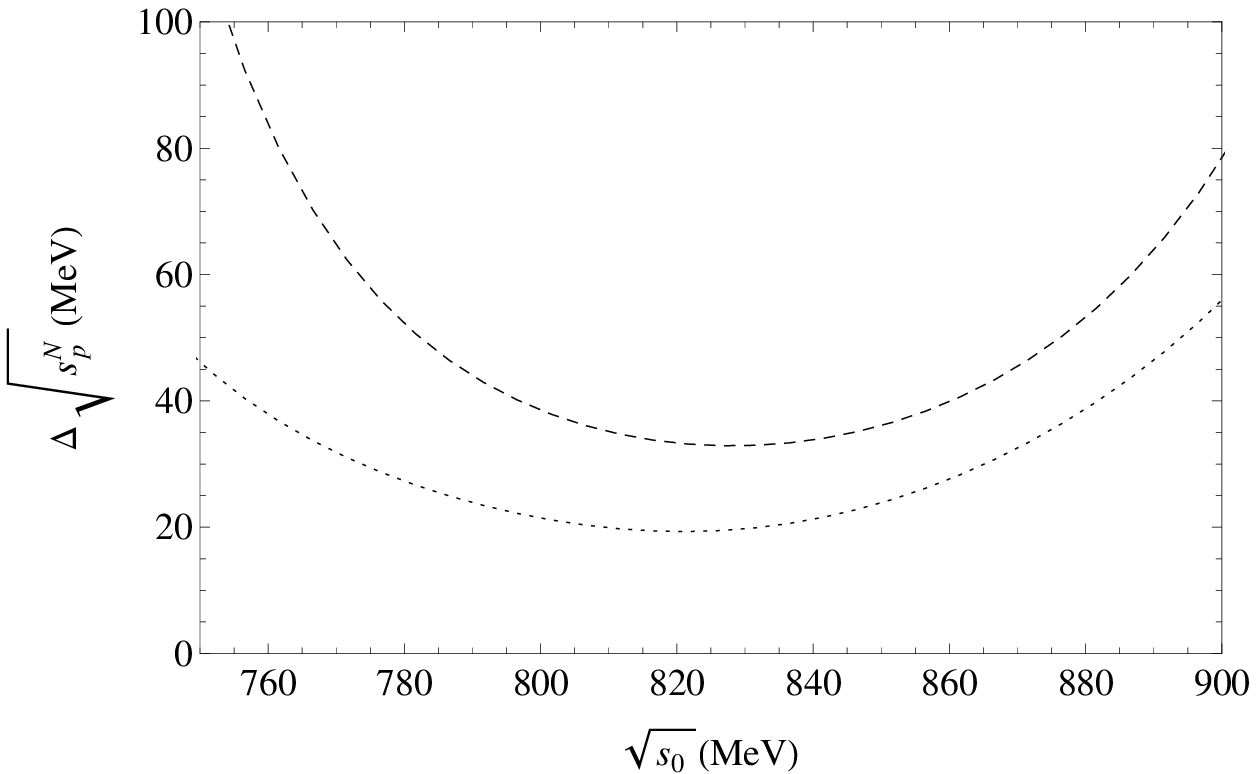}  }
\centerline{ \includegraphics[width=0.8\linewidth]{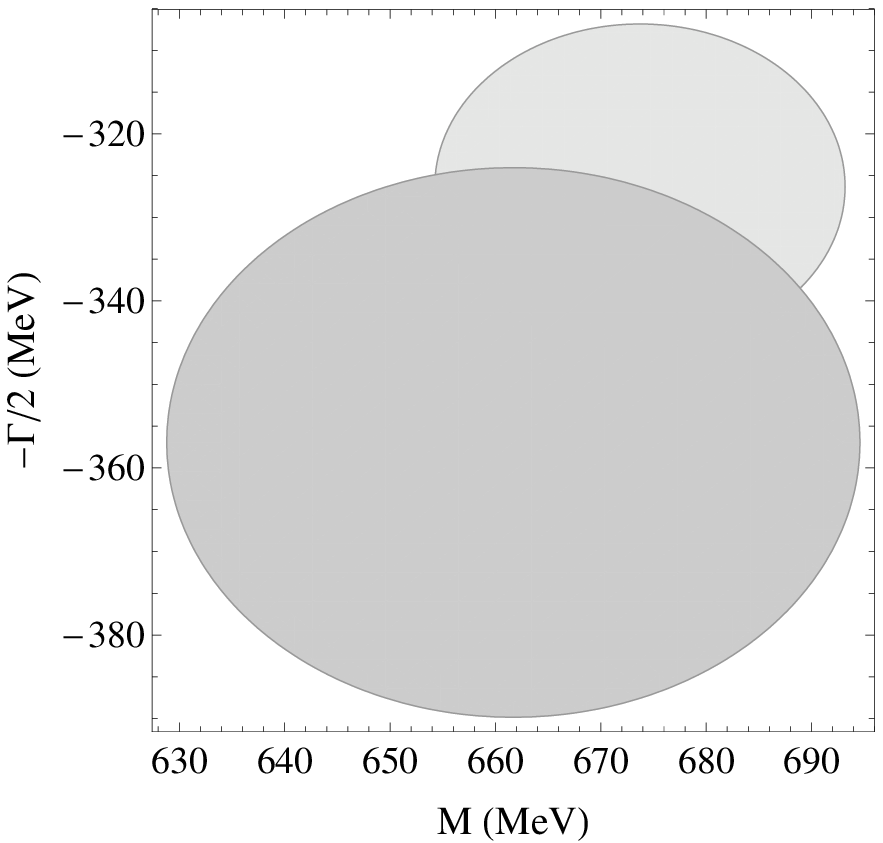} }
\caption{\rm \label{fig:trunkappa} 
Upper panel, uncertainty $\Delta \sqrt{s_p^N}$ for different values of $\sqrt{s_0}$ in the $\kappa$ pole determination for M=1.
We show dotted and dashed lines for 
N=2 and N=3 respectively. It is clear that the M=1 case does not converge as N increases. We have checked higher $N$ and there is no improvement.
Lower panel, theoretical uncertainty regions $\Delta \sqrt{s_p^N}$ for the best center $\sqrt{s_0}$ for M=1, where N=2 is plotted as the light gray region and N=3 as the gray region.
}
\end{figure}

The results for $M=1$
can be found in 
Fig.~\ref{fig:trunkappa}. In the upper panel 
 we show $\Delta\sqrt{s_p^N}$ 
for different values of $s_0$.
Note that the $N=3$ curve (dashed) is nowhere smaller then that of $N=2$ (dotted).
The smallest uncertainty for each $N$ is attained at $\sqrt{s_0}\sim 830\,$MeV, 
and we show in the lower panel how it translates into a 
truncation uncertainty for the pole position, which grows from $N=2$ (light gray circle)
to $N=3$ (darker gray circle). Note also that the central value of the 
darker circle lies well outside the lighter circle.
We have also calculated the $N=4,5$ cases 
and there is no evidence of convergence for $M=1$.
We thus conclude that considering the Pad\'e series with just one pole 
is not enough
to reproduce the analytic structure in the region relevant for such a deep pole.

\begin{figure}
\centering
\centerline{ \includegraphics[width=1\linewidth]{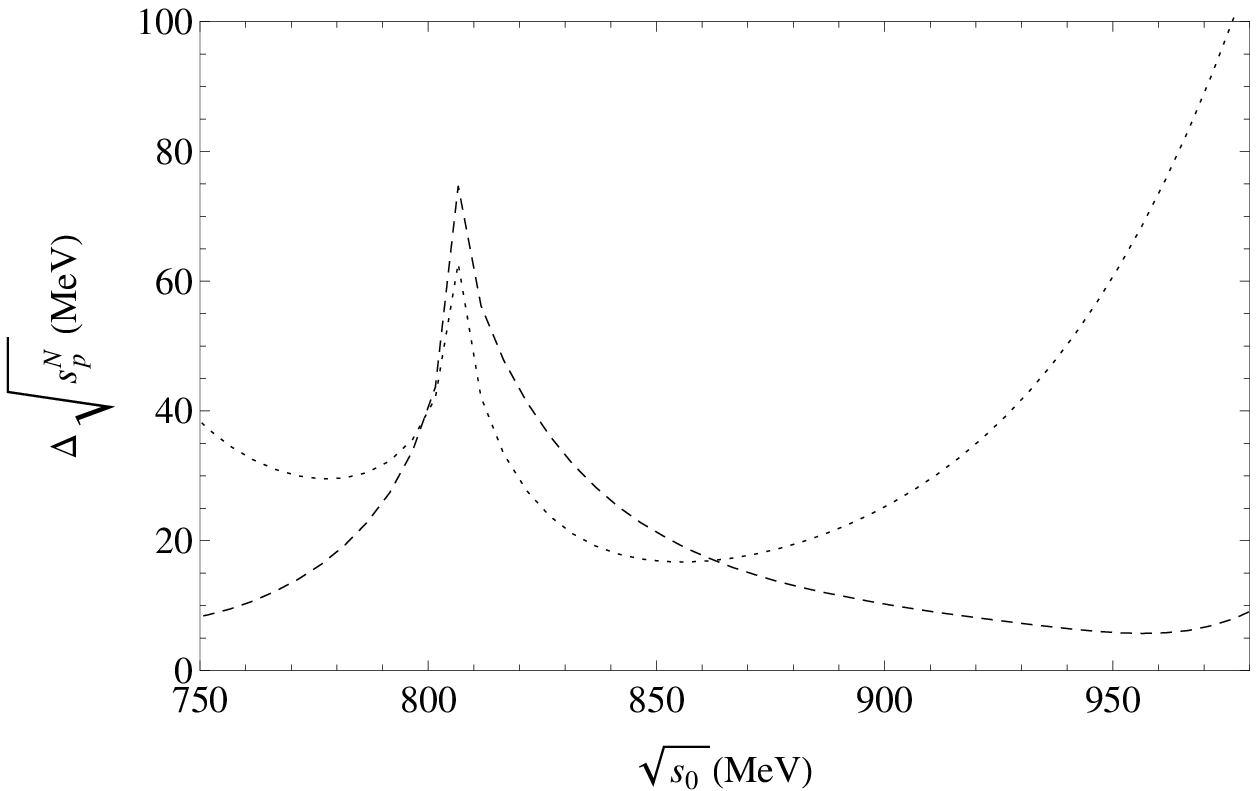} }
\centerline{ \includegraphics[width=0.8\linewidth]{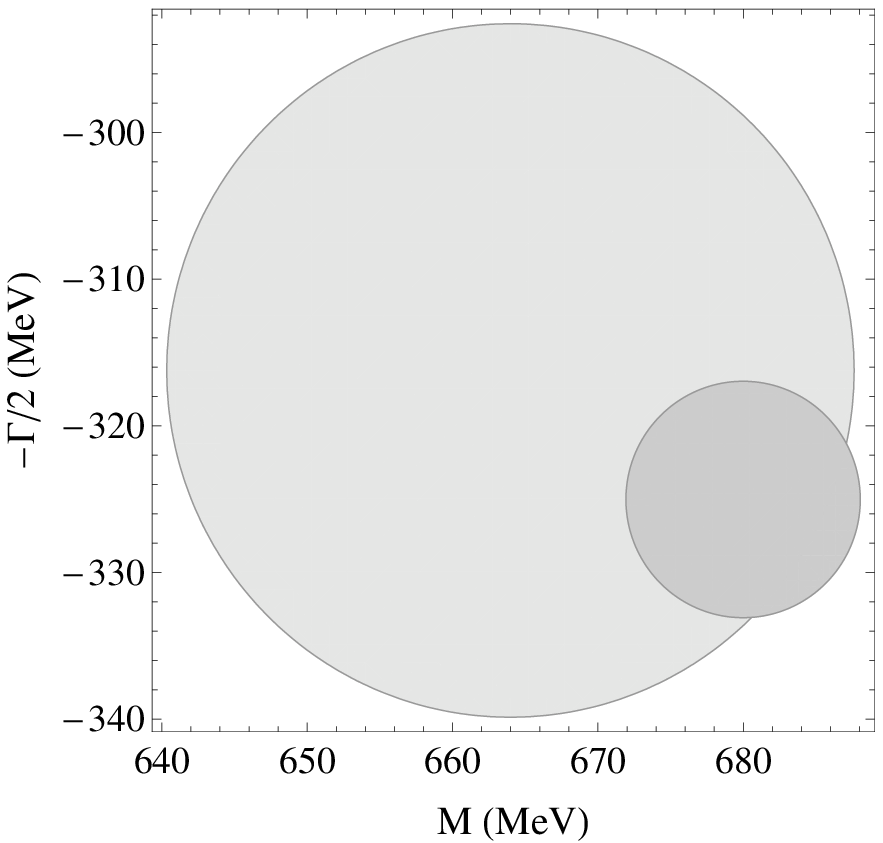}}
\caption{\rm \label{fig:Kappatrun} 
Upper panel, uncertainty $\Delta \sqrt{s_p^N}$ for different values of $\sqrt{s_0}$ in the $\kappa$ pole determination for M=2.
We show dotted and dashed lines for 
N=2 and N=3 respectively. 
Lower panel, theoretical uncertainty regions $\Delta \sqrt{s_p^N}$ for the best center $\sqrt{s_0}$ for M=2, where N=2 is plotted as the light gray region and N=3 as the gray region.
}
\end{figure}

We then show in Fig.~\ref{fig:Kappatrun} 
the results for the $M=2$ Pad\'e series, which has two poles. 
Once again, in the upper panel we show
$\Delta\sqrt{s_p^N}$, for different values of $s_0$, 
as dotted and dashed curves for $N=2$ and $N=3$, respectively.
Now we see that this truncation difference 
decreases drastically in several $s_0$ regions as $N$ increases.
Actually, already at $N=3$ it becomes smaller 
than the statistical uncertainties, with a minimum at 
$\sqrt{s_0}=950\,$MeV. Thus the $P^3_2$ pole 
will define our resonance values and $\Delta\sqrt{s_p^3}$ the theoretical uncertainty
$\Delta_{th}$ listed in Table~\ref{tab:kappap32}.
In the lower panel
we show the pole position and its minimum truncation uncertainty 
for $N=2,3$ as the light and dark gray areas, respectively. 
The other pole obtained for this sequence corresponds to the $\eta K$ threshold,
which is the nearest singularity to $s_0$.

Once a central value and a theoretical error for the pole position has been obtained, 
we add the statistical uncertainty in quadrature: 
\begin{equation}
\Delta_{s_p}=\sqrt{\Delta_{th}^2+\Delta_{stat}^2}.
\end{equation}
Recall that the statistical errors are obtained from a Monte Carlo 
Gaussian sampling of the parameters of the CFD parameterization 
within their uncertainties.
Statistical uncertainties dominate the quadrature,
since the theoretical error is the $\Delta\sqrt{s_p^N}$ for the $N$ when 
it becomes smaller
than the experimental one.  
In the case of the $K^*(800)$ this procedure leads to the results for the 
pole position and the coupling that are listed in the second column of 
Table~\ref{tab:kappap32}. The central value $(680\pm13)-i(325\pm7)\,$MeV
obtained with the Pad\'e approximant can now be compared with the 
pole position extracted analytically
from the CFD parameterization in $(680\pm15)-i(334\pm7.5)\,$MeV.
This illustrates the remarkable accuracy of the Pad\'e sequence
to extract resonance parameters and the soundness of our  method to estimate uncertainties.

\begin{table} 
\setlength{\tabcolsep}{3pt}
\caption{$K_0^*(800)$ pole results for the CFD and different parameterizations fitted to the CFD.
The uncertainty for $\sqrt{s_p}$ and $g$ include statistical and theoretical errors only. } 
\centering 
\begin{tabular}{cccc} 
\hline\hline  
\rule[-0.15cm]{0cm}{.15cm}  & CFD Pad\'e& Schenk Pad\'e& C-M Pad\'e \\ 
\hline
\rule[-0.15cm]{0cm}{.15cm} $\sqrt{s_p}$(MeV) & (680$\pm$13)-$i$(325$\pm$7) & (656)-$i$(283) & (673)-$i$(276) \\
\rule[-0.15cm]{0cm}{.15cm} $\Delta_{th}$(MeV)& 6 & 13 & 10 \\
\rule[-0.15cm]{0cm}{.15cm} $g$(GeV)          & 4.88$\pm$0.16 & 4.30 & 4.22 \\
\rule[-0.15cm]{0cm}{.15cm} $\Delta_{th}$(GeV)& 0.15 & 0.32 & 0.20 \\
\rule[-0.15cm]{0cm}{.15cm} $\sqrt{s_0}$(GeV) & 0.96 & 0.81 & 0.87 \\
\hline
\end{tabular} 
\label{tab:kappap32} 
\end{table}

In the third and fourth columns of Table~\ref{tab:kappap32} we also show the results 
obtained by following the same procedure with
the Schenk \cite{Schenk:1991xe} and Chew-Mandelstam (C-M) \cite{Chew:1960iv}
parameterizations already  used in \cite{Caprini:2016uxy}  fitted 
to the CFD curve. For each parameterization we choose its best $s_0$ value.
As explained above, although these parameterizations
fall within the uncertainties of the CFD in the real axis, they yield 
slightly different derivatives, which result in somewhat different values for the pole.
Thus, we take as our final central result for the $K_0^*(800)$ resonance
the average of these different parameterizations and 
consider the systematic uncertainty as explained in the introduction, 
combining it quadratically with the theoretical and statistical uncertainties.
We thus arrive to the final result for the $K_0^*(800)$ pole and coupling:

\begin{eqnarray}
\sqrt{s_{K_0^*(800)}}&=&(670\pm18)-i(295\pm28)\;{\rm MeV},\\
g_{K_0^*(800)}&=&4.47\pm0.40\;{\rm GeV}. \nonumber
\end{eqnarray}

\begin{figure}
\centering
\includegraphics[width=1\linewidth]{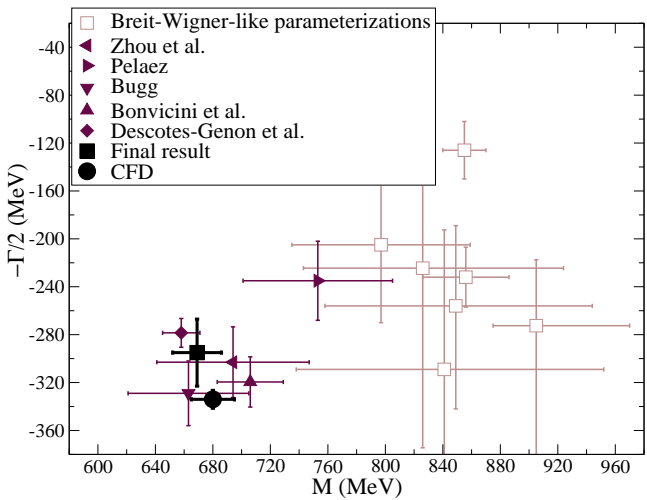}
\caption{\rm \label{fig:kappa} 
Final result for the $\kappa$ pole. Other references are taken from the RPP compilation \cite{PDG}, Descotes-Genon et al. \cite{DescotesGenon:2006uk}, Bonvicini et al. \cite{Bonvicini:2008jw}, D.Bugg \cite{Bugg:2003kj}, J.R.Pel\'aez \cite{GomezNicola:2001as}, Zhou et al. \cite{Zhou:2006wm}, and the Breit-Wigner parameterizations 
\cite{Ablikim:2010ab} listed in the RPP.}
\end{figure}

This result is shown in Fig.\ref{fig:kappa} together with the other references listed in the RPP for this resonance. Note that we have highlighted with solid symbols those poles coming from analytic or dispersive approaches, whereas mass and width values obtained from models using Breit-Wigner approximations are shown with empty squares.

\subsubsection{The $K_0^*(1430)$}

For the heavier $K_0^*(1430)$ resonance, the elastic formalism cannot be used, 
although the resonance is almost elastic, since its branching ratio 
to $\pi K$ is larger than $90\%$. In this case 
the CFD \cite{Pelaez:2016tgi} makes use of
and inelastic formalism parameterized through simple rational functions that fit the total 
phase and the modulus of the partial wave. 
Let us remark that, even for $\pi\pi$ scattering,  no partial-wave dispersion relations 
have been implemented up to more than 1.1 GeV,
since Roy and GKPY equations reach 1.1 GeV at most in their usual formulation.  Forward dispersion relations have been extended for $\pi\pi$ scattering up to 1420 MeV~\cite{GarciaMartin:2011cn} and for $K\pi$ up to 1600 MeV~\cite{Pelaez:2016tgi}, 
but they are not suitable for resonance pole extractions.
Therefore, lacking these rigorous dispersive methods to extract poles,
it is here where the Pad\'e technique yields more relevant results, providing a sound analytic
continuation 
to the next Riemann sheet.

\begin{figure}
\centering
\includegraphics[width=1\linewidth]{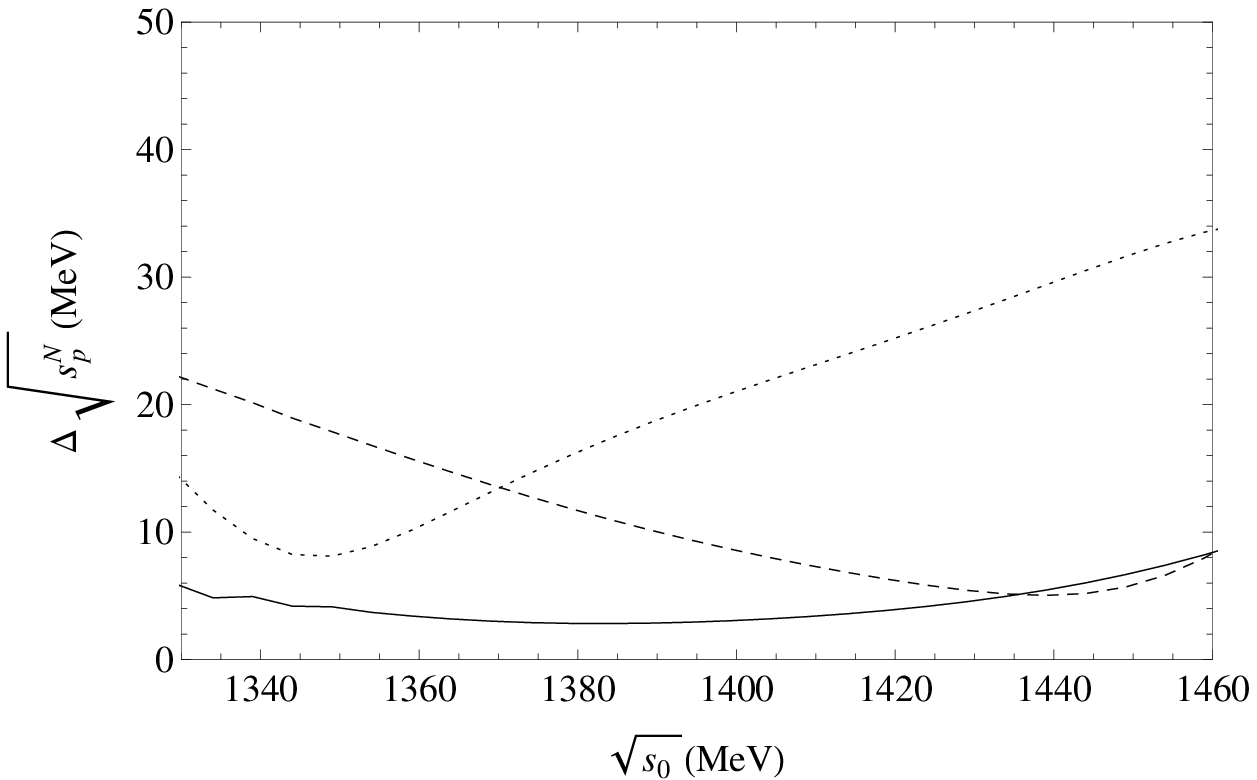}
\includegraphics[width=0.8\linewidth]{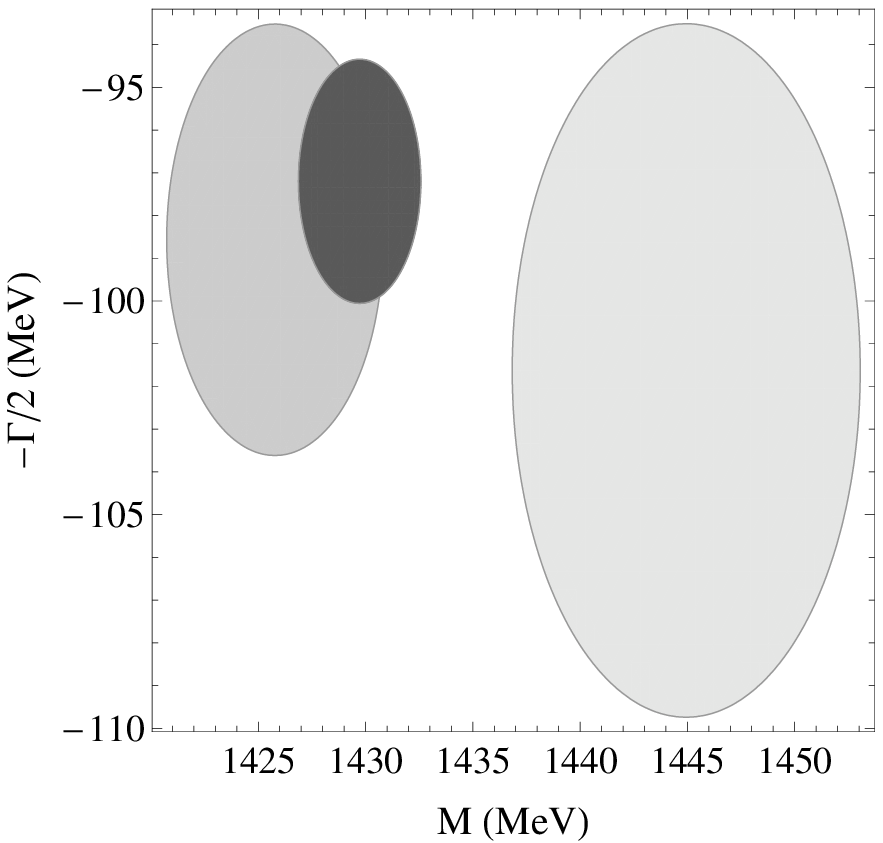}
\caption{\rm \label{fig:trunk0} 
Upper panel, uncertainty $\Delta \sqrt{s_p^N}$ for different values of $\sqrt{s_0}$ in the $K_0^*(1430)$ pole determination. The dotted, dashed and continuous
 lines correspond to the $N=2,3$ and 4 cases, respectively.
Lower panel, theoretical uncertainty 
regions $\Delta \sqrt{s_p^N}$ for the $K_0^*(1430)$ pole.
The light gray, gray and dark grey areas correspond to $N=2,3$ and 4.
}
\end{figure}

The convergence of the $P^N_1$ sequence, with just one pole, is fairly good this time because the
resonance is not as deep in the complex plane as the $K_0^*(800)$. 
In particular, the truncation errors, shown in the upper panel of Fig.~\ref{fig:trunk0}
decrease from $N=2$ to $4$
rather fast for $s_0$ within the $1350-1420$ MeV range. 
We obtain a minimum for the combined $\Delta_{s_p}$ error at $\sqrt{s_0}=1380$ MeV.
Once more,  the Pad\'e series has been truncated at $N=4$, where the theoretical error becomes smaller than the statistical one calculated 
from a Monte Carlo Gaussian sampling of the CFD parameters.
This truncation uncertainty translates into the light gray, gray and dark gray areas in 
the lower panel of Fig.~\ref{fig:trunk0}. The darker one gives our final 
central value and theoretical uncertainty, whose numerical values can be read in
the second column of
Table~\ref{tab:k0}. In addition, we have added in quadrature the statistical uncertainty
in the first line.

In that Table we also show the result of using a typical Breit-Wigner model, as done in most of the works
listed in the RPP, to fit the CFD parameterization. As it can be seen in the third column of Table~\ref{tab:k0}, this leads to a sizable
change in the width, but to almost an imperceptible variation of the mass. This is a source of systematic uncertainty due to model dependence.
Our final result is obtained by combining the three sources of uncertainty: theoretical, 
statistical, and systematic. We find
\begin{eqnarray}
\sqrt{s_{K_0^*(1430)}}&=&(1431\pm6)-i(110\pm19)\;{\rm MeV},\\
g_{K_0^*(1430)}&=&3.82\pm0.74\;{\rm GeV}. \nonumber
\end{eqnarray}

\begin{figure}
\centering
\includegraphics[width=1\linewidth]{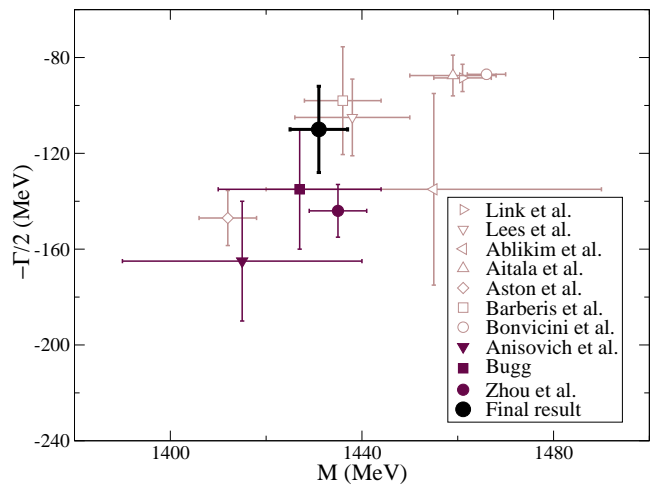}
\caption{\rm \label{fig:k0} 
Final result for the $K_0^*(1430)$ pole. Other results correspond to those
listed in the RPP compilation \cite{PDG}, Zhou et al. \cite{Zhou:2006wm}, D.Bugg \cite{Bugg:2003kj}, Anisovich et al. \cite{Anisovich:1997qp}, Bonvicini et al. \cite{Bonvicini:2008jw}, Barberis et al. \cite{Barberis:1998tv}, Aston et al. \cite{Aston:1987ir}, Aitala et al. \cite{Ablikim:2010ab}, Ablikim et al. \cite{Ablikim:2005kp}, Lees et al. \cite{Lees:2014iua}, Link et al. 
\cite{Pennington:2007se}.
}
\end{figure}

In Fig.~\ref{fig:k0} we have plotted this final result value as a black circle,
which compares rather well with the references listed in the RPP.

\begin{table} 
\caption{$K_0^*(1430)$ pole results for the CFD and different parameterizations fitted to the CFD.
The uncertainty for $\sqrt{s_p}$ and $g$ include statistical and theoretical errors only. } 
\centering 
\begin{tabular}{c c c} 
\hline\hline  
\rule[-0.15cm]{0cm}{.35cm}  & CFD Pad\'e & BW Pad\'e \\ 
\hline
\rule[-0.15cm]{0cm}{.35cm} $\sqrt{s_p}$(MeV) & (1430$\pm$5)-$i$(97$\pm$6) & (1431)-$i$(122) \\
\rule[-0.15cm]{0cm}{.35cm} $\Delta_{th}$(MeV)& 3 & 7 \\
\rule[-0.15cm]{0cm}{.35cm} $g$(GeV)          & 3.31$\pm$0.21 & 4.32 \\
\rule[-0.15cm]{0cm}{.35cm} $\Delta_{th}$(GeV)& 0.06 & 0.07 \\
\rule[-0.15cm]{0cm}{.35cm} $\sqrt{s_0}$(GeV) & 1.38 & 1.44 \\
\hline
\end{tabular} 
\label{tab:k0}
\end{table}

\subsection{Vector resonances}

Let us now discuss the vector resonances that
 appear in $K\pi$ scattering below 1.8 GeV. These are the $K^*(892)$
and the $K_1^*(1410)$, both of them with isospin $1/2$.

\subsubsection{The $K^*(892)$}

\begin{figure}
\centering
\includegraphics[width=1\linewidth]{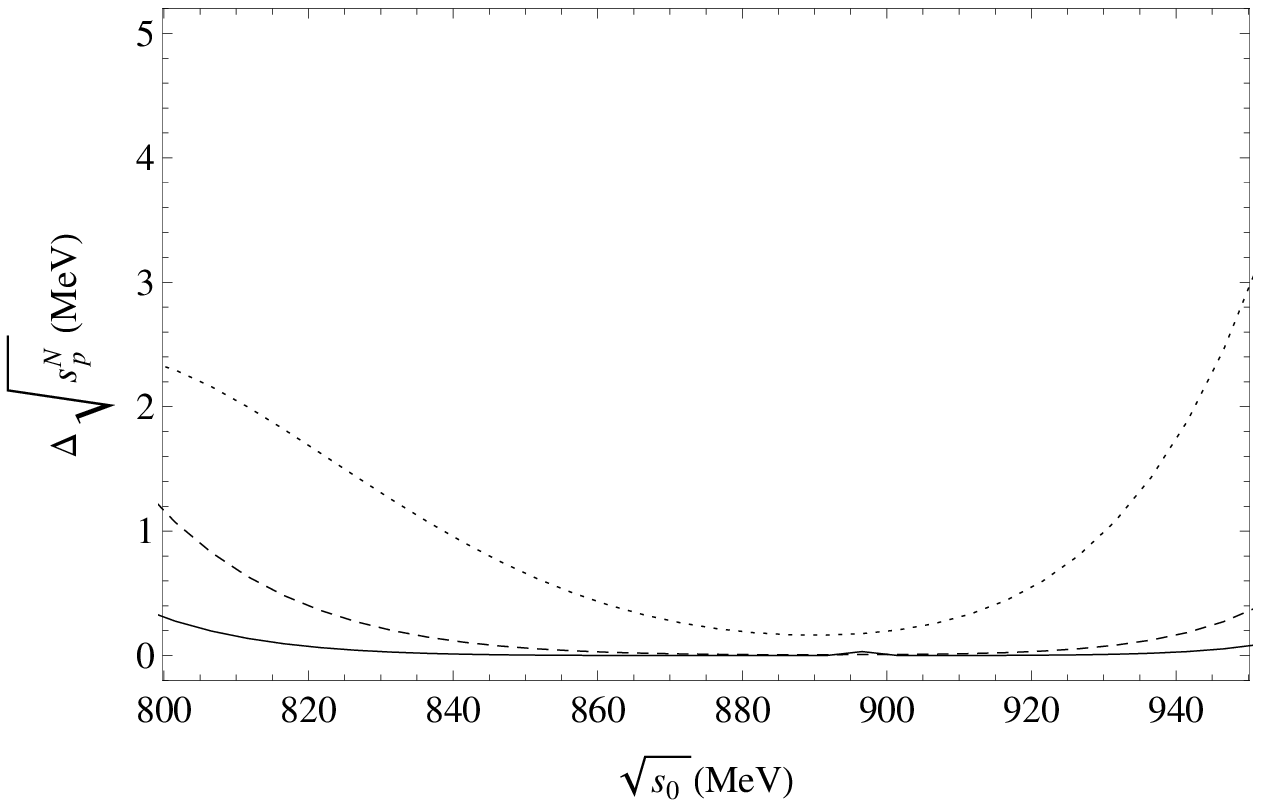}
\includegraphics[width=0.8\linewidth]{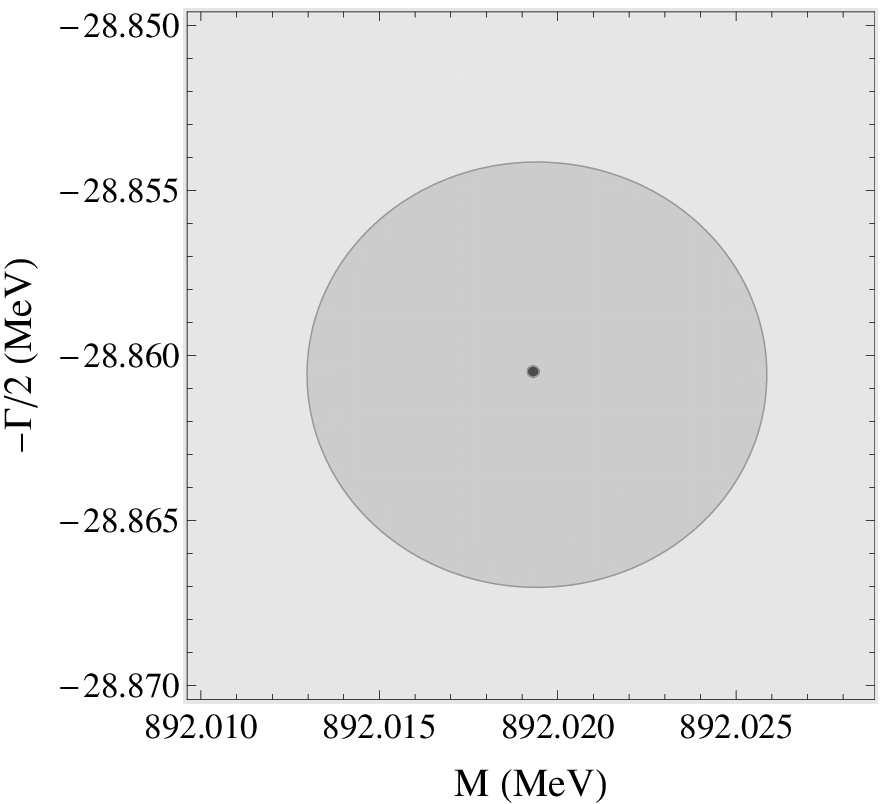}
\caption{\rm \label{fig:trunkstar} 
Upper panel: $\Delta \sqrt{s_p^N}$ in the $K^*(892)$ pole determination
for different values of $\sqrt{s_0}$ using the $P_1^N$ sequence.
 The dotted, dashed and continuous lines correspond to $N=2,3$ and 4, respectively. 
Lower panel: 
theoretical uncertainty regions $\Delta \sqrt{s_p^N}$ for the
$K^*(892)$ pole. The light gray and gray areas correspond to $N=2,3$, 
whereas the $N=4$ case corresponds to the tiny dark gray spot in the center,
since the theoretical uncertainty becomes negligible.
}
\end{figure}

The lightest one is the $K^*(892)$, which is elastic for all means and purposes. 
It is also very narrow compared to the $K_0^*(800)$ and therefore
much closer to the real axis and well isolated from other analytic structures.
Hence, as can be seen in the upper panel of Fig.~\ref{fig:trunkstar},
the $P^N_1$ sequence, with just one pole, converges very rapidly.
Actually, we estimate our theoretical error 
from $N=4$ since it is when the truncation error becomes 
negligible  compared to 
the statistical one (four orders of magnitude smaller), obtained as usual from a Monte Carlo Gaussian sampling of the CFD parameters. The $\Delta_{s_p}$ error is minimized for $\sqrt{s_0}=890$ MeV, but 
the outcome is remarkably similar within the $870-910$ MeV range.
In the lower panel of Fig.~\ref{fig:trunkstar} 
we see that the theoretical uncertainty on the pole becomes extremely small
and that he convergence is remarkable, with the central value being almost the same 
from $N=2$ to $N=4$.

Concerning the systematic uncertainty due to the use of other models to fit the same data,
we have found that the result,
if we consider a Breit-Wigner model fitted to the CFD values, 
differs by less than $1$ MeV. However, it is worth noting
that when fitting a Breit-Wigner to the CFD result,
the sequence of Pad\'e approximants with just one pole converges rather poorly. 
We have also tried other conformal parameterizations with different centers.
In any case, by changing the model, the systematic
uncertainty is smaller than the statistical  error,
which dominates the uncertainty in our final result.

The final result for the $K^*(892)$ parameters is shown in Table~\ref{tab:kstar}. 
This result may appear incompatible 
with the determinations in the RPP.  There are several reasons for this: first, because 
in the RPP only Breit-Wigner (BW) parameters are given and then 
$s_p=M_{BW}^2-i M_{BW}\Gamma_{BW}$,
so that $\re \sqrt{s_p}$ is not exactly $M_{BW}$ and $\im \sqrt{s_p}$ is not exactly $\Gamma_{BW}/2$. Taking these different definitions into account improves slightly the agreement.
 Second, there is the issue of using an
isospin conserving formalism when extracting the CFD parameterization 
and when measuring the data, so that our
resonances do not correspond to the charged nor the neutral cases. Therefore, when comparing to 
the resonances observed in a charged or neutral channels, which are the ones listed in the PDG,
a difference of about $\pm 2\,$MeV
is expected to arise. However, our pole is to be understood as the 
pole in the isospin conserving limit. Note that this distinction between
charges and neutral resonances is not done in the RPP for other resonances.
Moreover, there is a third reason,
which is that the BW extractions of resonance parameters are usually obtained from a fit to the amplitude in a limited region or assuming the existence of a certain background from other regions or resonances. In contrast, here the whole elastic 
region is described with the CFD amplitude, 
thus, we are giving the pole of the whole amplitude.
In general we do not think that obtaining this particular resonance 
from scattering data is competitive with the determinations from other reactions, which much better data and statistics.

\begin{table} 
\caption{$K^*(892)$ pole results.
The uncertainty for $\sqrt{s_p}$ and $g$ include statistical and theoretical errors only. } 
\centering 
\begin{tabular}{c c} 
\hline\hline  
\rule[-0.15cm]{0cm}{.35cm}  & CFD Pad\'e \\ 
\hline
\rule[-0.15cm]{0cm}{.35cm} $\sqrt{s_p}$(MeV) & (892$\pm$1)-$i$(29$\pm$1) \\
\rule[-0.15cm]{0cm}{.35cm} $\Delta_{th}$(MeV)& 1$\cdot{10^{-4}}$ \\
\rule[-0.15cm]{0cm}{.35cm} $g$          & 6.1$\pm$0.1 \\
\rule[-0.15cm]{0cm}{.35cm} $\Delta_{th}$& $\simeq$0 \\
\rule[-0.15cm]{0cm}{.35cm} $\sqrt{s_0}$(GeV) & 0.89 \\
\hline
\end{tabular} 
\label{tab:kstar} 
\end{table}

\subsubsection{The $K_1^*(1410)$}

Let us now turn to the $K_1^*(1410)$, which 
cannot be considered elastic
and has a rather small $7\%$ branching fraction to  $K\pi$.
Still, we will be able to obtain its pole, as we did for the 
$K_0^*(1430)$, since Pad\'e approximants also
provide the analytic continuation to the continuous Riemann 
sheet of the
partial waves in the inelastic region. Once more it is enough to compute derivatives from the vector
partial-wave CFD parameterization in \cite{Pelaez:2016tgi}. 

\begin{figure}
\centering
\includegraphics[width=1\linewidth]{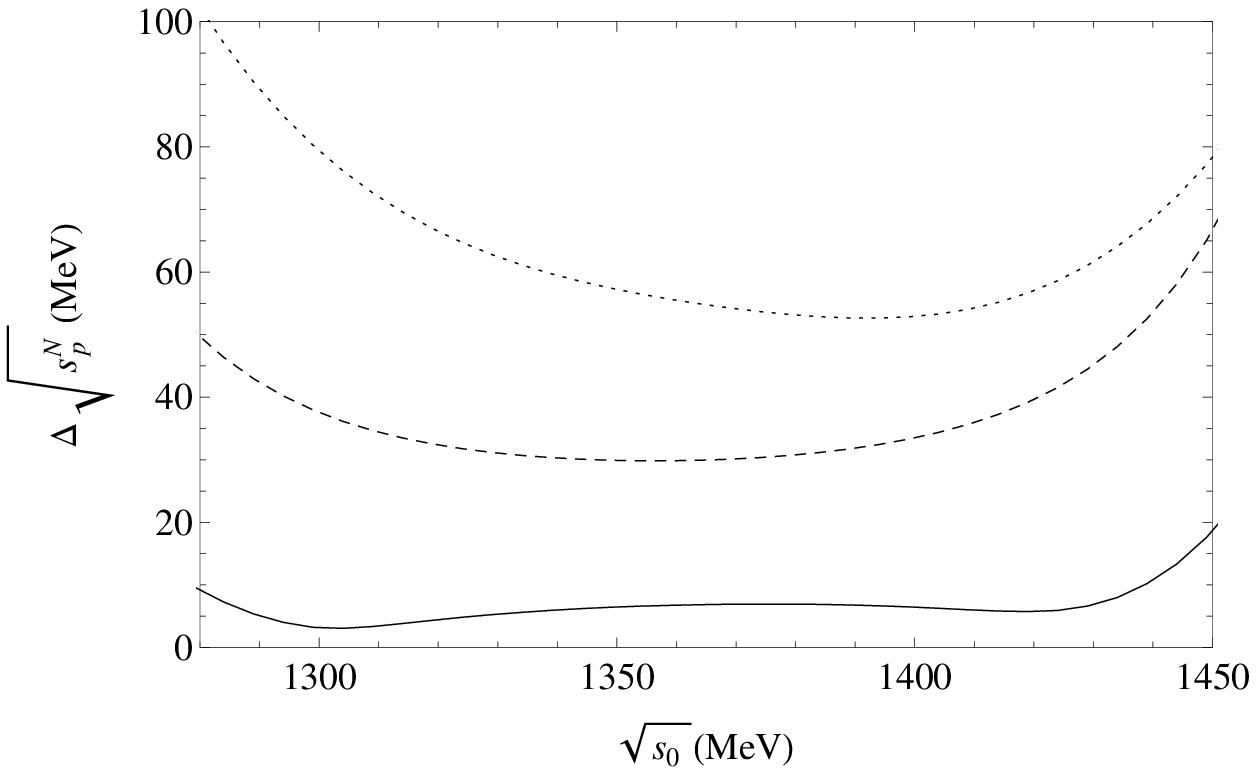}
\includegraphics[width=0.8\linewidth]{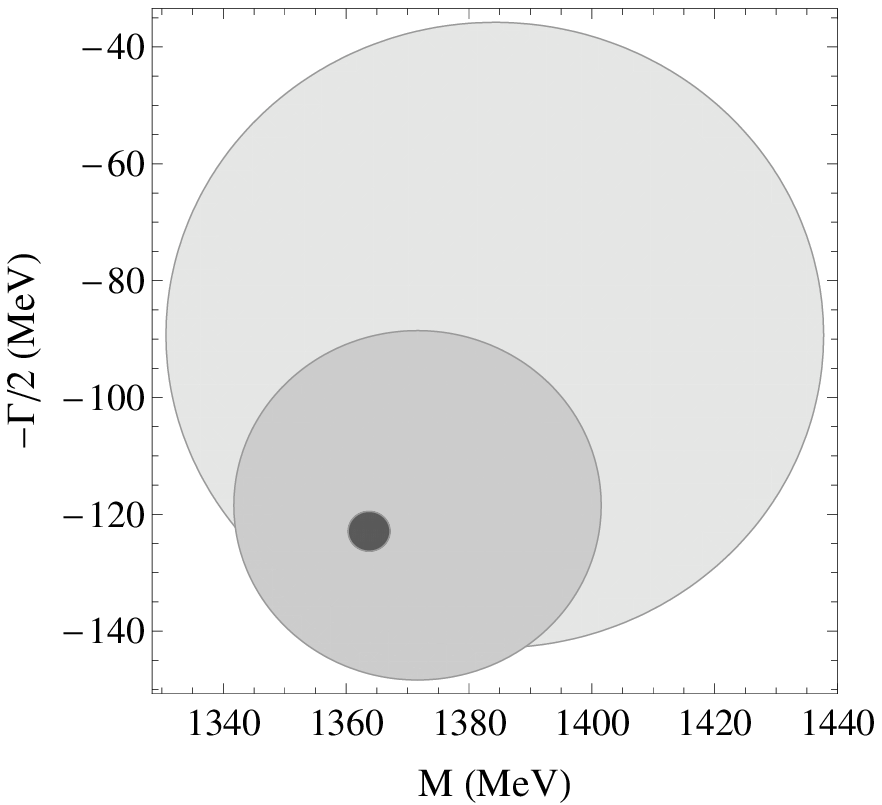}
\caption{\rm \label{fig:trunk1star} 
Upper panel: $\Delta \sqrt{s_p^N}$ in the $K_1^*(1410)$ pole determination
for different values of $\sqrt{s_0}$ using the $P_1^N$ sequence.
 The dotted, dashed and continuous lines correspond to $N=2,3$ and 4, respectively. 
Lower panel: 
theoretical uncertainty regions $\Delta \sqrt{s_p^N}$ for the
$K_1^*(1410)$ pole. The light gray, gray and dark gray areas correspond 
to $N=2,3$ and 4.
}
\end{figure}

The theoretical convergence 
is really fast as can be observed in Fig.~\ref{fig:trunk1star}.
The theoretical error is small in the range $1280-1450\,$MeV, 
with a minimum for the total error located at $\sqrt{s_0}=1304$ MeV. In this case the theoretical uncertainty becomes much smaller than the statistical one at $N=4$. Partly, this is due to the fact that in this energy region
there are two conflicting experiments and this leads to large uncertainties in the CFD parameterization. As a consequence, what  we call ``statistical'' uncertainties dominate 
the final result for this resonance.

As with other resonances, we have also fitted other parameterizations to the CFD data
to estimate the systematic uncertainty when calculating the derivatives at one given energy.
In Table~\ref{tab:k1star} we show the results when calculating the derivatives with a BW formalism, which is the one used by all the determinations quoted in the RPP \cite{PDG}.
For the final central value we thus take the average over these two determinations
and we evaluate our final error as the quadrature between statistical, theoretical, and systematic uncertainties as:
\begin{eqnarray}
\sqrt{s_{K_1^*(1410)}}&=&(1368^{+38}_{-38})-i(106^{+48}_{-59})\;{\rm MeV},\\
g_{K_1^*(1410)}&=&1.89^{+1.77}_{-1.34}. \nonumber
\end{eqnarray}

\begin{figure}
\centering
\includegraphics[width=1\linewidth]{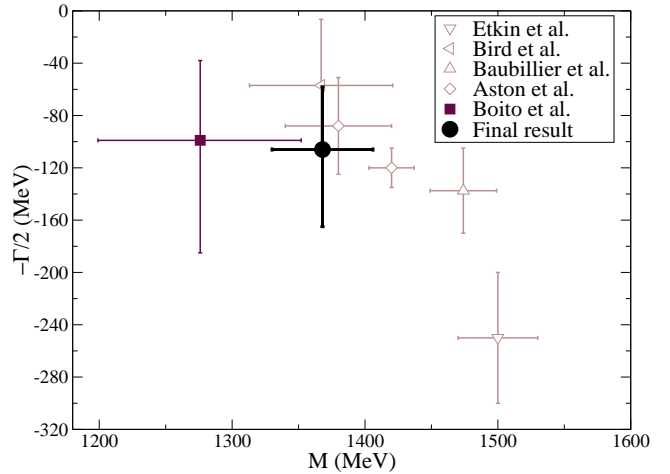}
\caption{\rm \label{fig:k1star} 
Final result for the $K_1^*(1410)$ pole. Other references are taken from the RPP compilation \cite{PDG}, Boito et al. \cite{Boito:2008fq}, Aston et al. \cite{Aston:1987ir}, Baubillier et al. \cite{Baubillier:1982eb}, Bird et al. \cite{Bird:1988qp}, Etkin et al. \cite{Etkin:1980me}.
}
\end{figure}

This might look less precise than the averaged result 
of $M=1414\pm15\,$MeV and $\Gamma/2=116\pm10.5$
given in the RPP \cite{PDG}, but this is because this average 
is dominated by a measurement of the LASS experiment on $K^-p\rightarrow\bar K^0\pi^+\pi^-n$ 
\cite{Aston:1986rm} using a BW parameterization with simple backgrounds. 
It is not evident the systematic effect due to these simple backgrounds.
When using the 
$K\pi$ scattering data obtained later by the same experiment \cite{Aston:1987ir}
one obtains $M=1380\pm21\pm19$ and $\Gamma/2=88\pm26\pm11$, 
very similar to our extraction, but based only on a BW formalism 
and without taking into account the conflicting data of Estabrooks 
et al. \cite{Estabrooks:1977xe} in this region. In this sense 
we think our result is more robust
and confirms 
the parameters of this resonance without using a specific BW functional form, nor 
assuming any particular background.
In Fig.~\ref{fig:k1star} we show how our final result compares to all other results listed in the RPP.
It can be seen that the results are rather consistent with the exception of that of Etkin et al. \cite{Etkin:1980me}.

\begin{table} 
\caption{$K_1^*(1410)$ pole results for the CFD and different parameterizations fitted to the CFD.
The uncertainty for $\sqrt{s_p}$ and $g$ include statistical and theoretical errors only. } 
\centering 
\begin{tabular}{c c c} 
\hline\hline  
\rule[-0.15cm]{0cm}{.35cm}  & CFD Pad\'e & BW Pad\'e \\ 
\hline
\rule[-0.15cm]{0cm}{.35cm} $\sqrt{s_p}$(MeV) & $(1362^{+37}_{-37})-$i$(123^{+41}_{-54})$ & $(1374)-$i$(88)$ \\
\rule[-0.15cm]{0cm}{.35cm} $\Delta_{th}$(MeV)& 3 & 0.7 \\
\rule[-0.15cm]{0cm}{.35cm} $g$         & $2.41^{+1.60}_{-1.11}$ & 1.36 \\
\rule[-0.15cm]{0cm}{.35cm} $\Delta_{th}$& 0.04 & 0.007 \\
\rule[-0.15cm]{0cm}{.35cm} $\sqrt{s_0}$(GeV) & 1.30 & 1.38 \\
\hline
\end{tabular} 
\label{tab:k1star} 
\end{table}

\subsection{Tensor resonances}

In practice, once we reach 1.3 GeV all available channels have some measured inelasticity.
Since all resonances with $J=2$ or higher angular 
momentum waves are above this energy, we use the inelastic CFD parameterization 
of \cite{Pelaez:2016tgi} 
and the fact that the Pad\'e approximants perform the analytical 
continuation directly to the continuous Riemann sheet. 
We describe next how
we extract the parameters of the $K_2^*(1430)$ and $K_3^*(1780)$ resonances, which have $J=2$ and 3
 respectively.

\subsubsection{The $K_2^*(1430)$}

This resonance appears in $K\pi$ scattering with angular momentum $2$ and
isospin $1/2$ and shows a nice Breit-Wigner-like shape. Its
branching ratio to $K\pi$ is $50\%$, the other relevant channels 
being $K^*(892) \pi$, $K^*(892) \pi \pi$ and $K \rho$.

\begin{figure}
\centering
\includegraphics[width=1\linewidth]{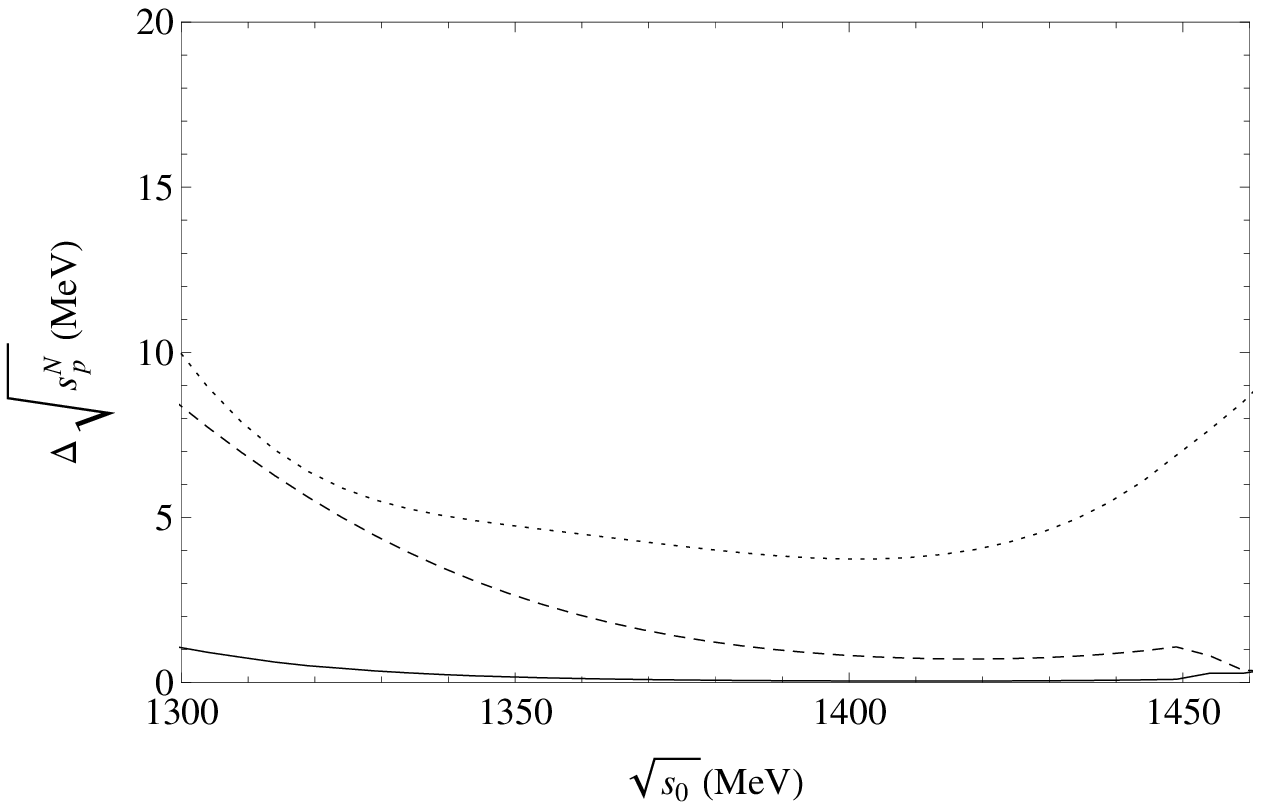}
\includegraphics[width=0.8\linewidth]{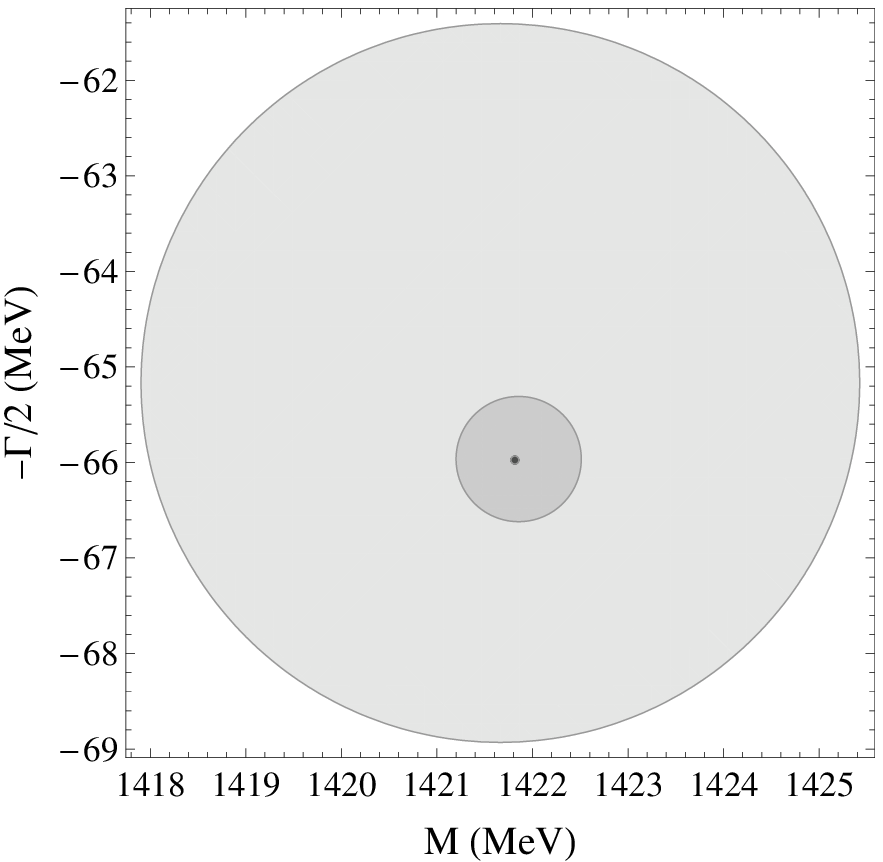}
\caption{\rm \label{fig:trunk2star} 
Upper panel: $\Delta \sqrt{s_p^N}$ in the $K_2^*(1430)$ pole determination
for different values of $\sqrt{s_0}$ using the $P_1^N$ sequence.
 The dotted, dashed and continuous lines correspond to $N=2,3$ and 4, respectively. 
Lower panel: 
theoretical uncertainty regions $\Delta \sqrt{s_p^N}$ for the
$K_2^*(1430)$ pole. The light gray, gray and dark gray areas correspond 
to $N=2,3$ and 4.
}
\end{figure}

Since there is a well isolated pole, we can use the $M=1$ Pad\'e sequence with one pole.
The upper panel of Fig.~\ref{fig:trunk2star}  shows how the sequence converges rapidly 
and for $N=4$ the truncation
uncertainty is completely negligible, having a minimum at $\sqrt{s_0}=1410$ MeV.
In the lower panel we see that the area
covered by the $N=4$ Pad\'e, almost becomes a point and that the
central value of the pole position is very stable.
The parameters of the resonance thus obtained are listed in Table~\ref{tab:k2star}.

\begin{figure}
\centering
\includegraphics[width=1\linewidth]{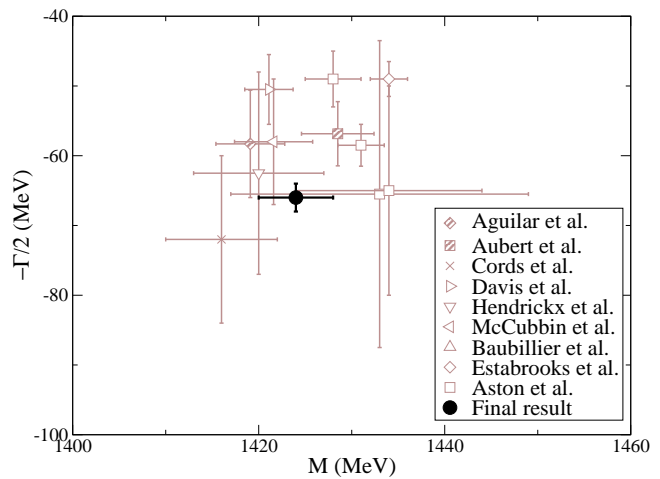}
\caption{\rm \label{fig:k2star} 
Final result for the $K_2^*(1430)$ pole. The following references
are taken from the RPP compilation \cite{PDG}, Aston et al. \cite{Aston:1987ir}, Estabrooks et al. \cite{Estabrooks:1977xe}, Baubillier et al. \cite{Baubillier:1982eb}, Mccubbin et al. \cite{McCubbin:1974yy}, Hendrickx et al. \cite{Hendrickx:1976ha}, Davis et al. \cite{Davis:1969sf}, Cords et al. \cite{Cords:1972hh}, Aubert et al. \cite{Aubert:2007ur}, Aguilar et al. \cite{AguilarBenitez:1972bw}.
}
\end{figure}

\begin{table} 
\caption{$K_2^*(1430)$ pole results for the CFD and different parameterizations fitted to the CFD.
The uncertainty for $\sqrt{s_p}$ and $g$ include statistical and theoretical errors only. } 
\centering 
\begin{tabular}{c c c} 
\hline\hline  
\rule[-0.15cm]{0cm}{.35cm}  & CFD Pad\'e & BW Pad\'e \\ 
\hline
\rule[-0.15cm]{0cm}{.35cm} $\sqrt{s_p}$(MeV) & (1422$\pm$3)-$i$(66$\pm$2) & (1427)-$i$(66) \\
\rule[-0.15cm]{0cm}{.35cm} $\Delta_{th}$(MeV)& 0.04 & 0.01 \\
\rule[-0.15cm]{0cm}{.35cm} $g$(GeV)$^{-1}$          & 3.37$\pm$0.07 & 3.08 \\
\rule[-0.15cm]{0cm}{.35cm} $\Delta_{th}$(GeV)$^{-1}$ & 0.001 & $3\times 10^{-5}$ \\
\rule[-0.15cm]{0cm}{.35cm} $\sqrt{s_0}$(GeV) & 1.41 & 1.51 \\
\hline
\end{tabular} 
\label{tab:k2star} 
\end{table}
As done with other resonances, we have tried calculating the derivatives needed for the Pad\'e approximants by means of other parameterizations fitted to the CFD results.
In particular we show in  Table~\ref{tab:k2star} the result when using a Breit-Wigner formula fitted to the CFD and then the Pad\'e approximants to extract the pole.
The difference is rather small, but we have taken the average with the CFD result obtained with Pad\'es and added the systematic uncertainty as explained in the introduction, yielding
our final result:
\begin{eqnarray}
\sqrt{s_{K_2^*(1430)}}&=&(1424\pm4)-i(66\pm2)\;{\rm MeV},\\
g_{K_2^*(1430)}&=&3.23\pm0.22\;{\rm GeV}^{-1}, \nonumber
\end{eqnarray}
which, as can be seen in Fig.~\ref{fig:k2star} is in good agreement with other 
determinations quoted in the RPP \cite{PDG}. The RPP average is dominated by the work of LASS 
\cite{Aston:1986rm,Aston:1987ir}, which use BW formalisms and simple backgrounds.
Our result has a relatively small
uncertainty despite including estimates of systematic error, both in the pole extraction and the data, and
avoiding the use of backgrounds or other assumptions in the pole extraction.

\subsubsection{The $K_3^*(1780)$}

The heaviest strange resonance that can be studied using the CFD parameterizations is the $K_3^*(1780)$, which appears in the $F$-wave with isospin 1/2. 
Let us note that the $K_3^*(1780)$ has a branching ratio to $\pi K$ of $20\%$, with the other 3 
relevant channels being $K \rho$, $K^*(892)$ and $K \eta$. 
First of all, let us remark that its mass lies beyond 1600 MeV, 
the energy up to which the CFD parameterization satisfies well the Forward Dispersion Relations.
Nevertheless, as explained in \cite{Pelaez:2016tgi}, this is most likely due 
to the data in other waves, since imposing FDRs up to higher energies demands deviations 
from the $D$-wave data, for instance, but the $F$-wave barely changes from an unconstrained fit
up to larger energies. Thus we feel confident our method can be applied to this resonance.

\begin{figure}
\centering
\includegraphics[width=1\linewidth]{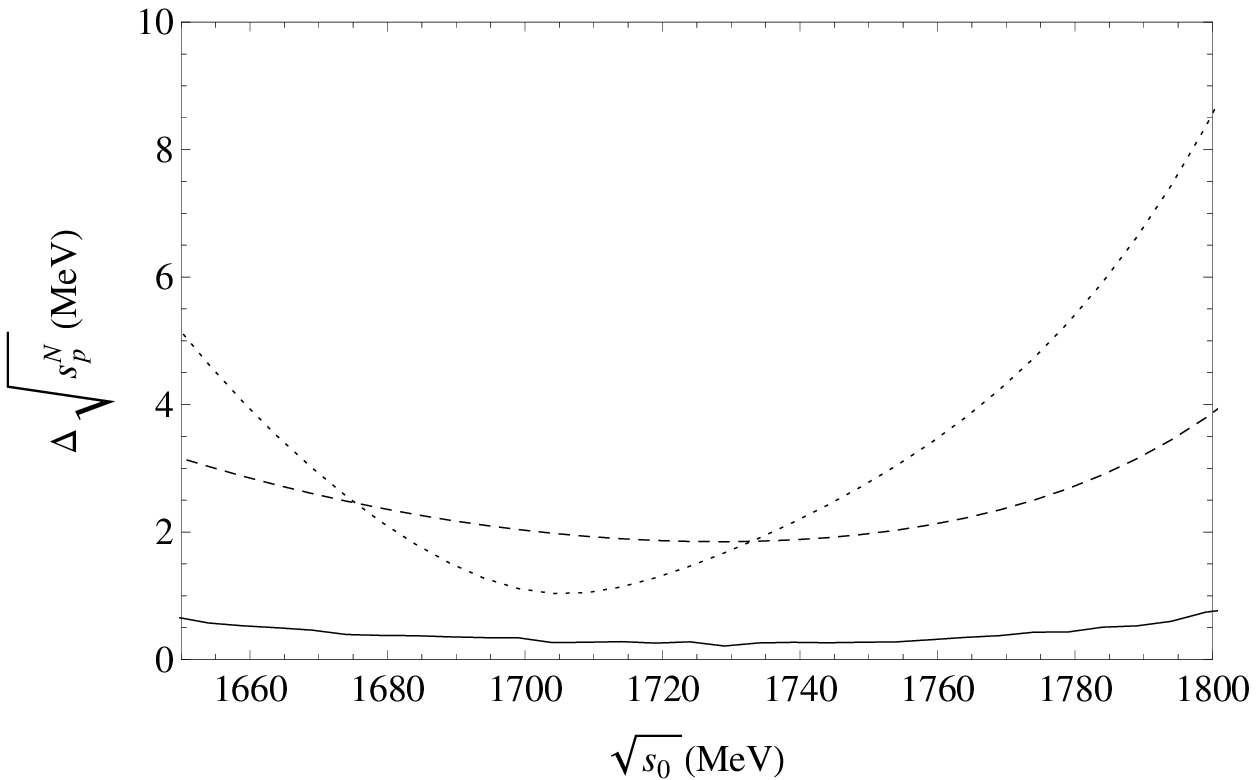}
\includegraphics[width=0.8\linewidth]{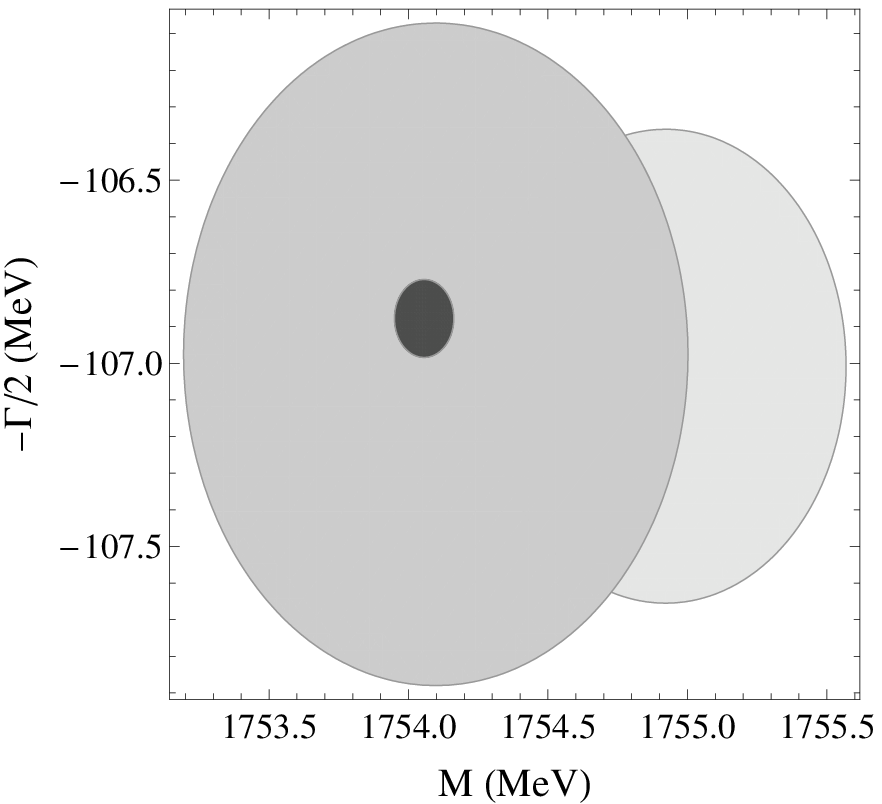}
\caption{\rm \label{fig:trunk3} 
Upper panel: $\Delta \sqrt{s_p^N}$ in the $K_3^*(1780)$ pole determination
for different values of $\sqrt{s_0}$ using the $P_1^N$ sequence.
 The dotted, dashed and continuous lines correspond to $N=2,3$ and 4, respectively. 
Lower panel: 
theoretical uncertainty regions $\Delta \sqrt{s_p^N}$ for the
$K_3^*(1780)$ pole. The light gray, gray and dark gray areas correspond 
to $N=2,3$ and 4.
}
\end{figure}

The $K_3^*(1780)$ is well isolated from contributions from other singularities and we
can use the Pad\'e sequence with just one pole.
As usual, we show in Fig.~\ref{fig:trunk3} the convergence of the sequence
which has a very small truncation error for $N=4$. 
Actually, it is about two orders of magnitude smaller than the statistical one,
as seen in Table~\ref{fig:trunk3}. As seen in the lower panel of that figure, the central value barely changes with $N$ (Note the small scale of the axis).

\begin{figure}
\centering
\includegraphics[width=1\linewidth]{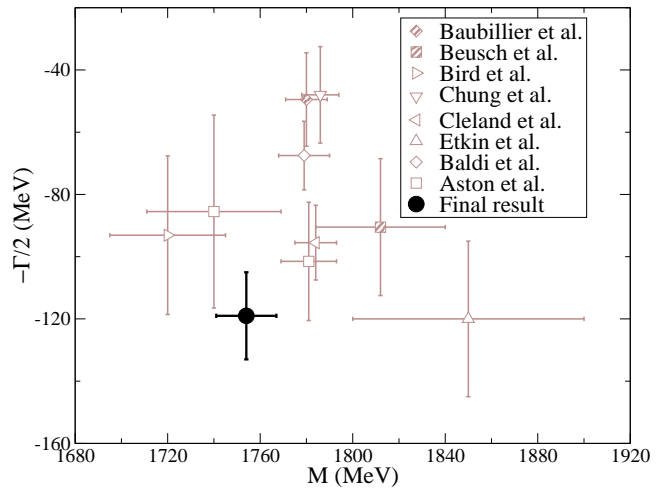}
\caption{\rm \label{fig:k3} 
Final result for the $K_3^*(1780)$ pole determination. We also show a  are taken from the PDG \cite{PDG}, Aston et al. \cite{Aston:1987ir}, Baldi et al. \cite{Baldi:1976ua}, Etkin et al. \cite{Etkin:1980me}, Cleland et al. \cite{Cleland:1982td}, Chung et al. \cite{Chung:1977ji}, Bird et al. \cite{Bird:1988qp}, Beusch et al. \cite{Konigs:1978at}, Baubillier et al. \cite{Baubillier:1982eb}.
}
\end{figure}

Once again we have tried to estimate the uncertainty due to calculating the derivatives of the amplitude with different parameterizations, but the differences are rather small.
In Table~\ref{fig:trunk3} we show the pole extracted with the Pad\'e method if, instead of using the CFD parameterizations, we use a BW fit to the  CFD. 
The mass and width barely change but the coupling is slightly different,
changing by less than the statistical uncertainty. We thus take the average and enlarge the uncertainty with 
a systematic error combined as explained in the introduction to this section.
Our result is:
\begin{eqnarray}
\sqrt{s_{K_3^*(1780)}}&=&(1754\pm13)-i(119\pm14)\;{\rm MeV},\\
g_{K_3^*(1780)}&=&1.28\pm0.14\;{\rm GeV}^{-2}, \nonumber
\end{eqnarray}
which, as seen in Fig.~\ref{fig:k3} is compatible with the 
results quoted in the RPP \cite{PDG}.  It should be noted that 
our uncertainties are only slightly larger than the RPP average, which is dominated by the result of 
Aston et al. \cite{Aston:1987ir}, but here we do not make a particular assumption for the functional form or a background in the amplitude.

\begin{table} 
\caption{$K_3^*(1780)$ pole results for the CFD and different parameterizations fitted to the CFD.
The uncertainty for $\sqrt{s_p}$ and $g$ include statistical and theoretical errors only. } 
\centering 
\begin{tabular}{c c c} 
\hline\hline  
\rule[-0.15cm]{0cm}{.35cm}  & CFD Pad\'e & BW Pad\'e \\ 
\hline
\rule[-0.15cm]{0cm}{.35cm} $\sqrt{s_p}$(MeV) & (1753$\pm$13)-$i$(119$\pm$14) & (1755)-$i$(118) \\
\rule[-0.15cm]{0cm}{.35cm} $\Delta_{th}$(MeV)& 0.3 & 4.3 \\
\rule[-0.15cm]{0cm}{.35cm} $g$(GeV)$^{-2}$           & 1.32$\pm$0.13 & 1.23 \\
\rule[-0.15cm]{0cm}{.35cm} $\Delta_{th}$(GeV)$^{-2}$ & 0.003 & 0.03 \\
\rule[-0.15cm]{0cm}{.35cm} $\sqrt{s_0}$(GeV) & 1.73 & 1.76 \\
\hline
\end{tabular} 
\label{tab:k3star} 
\end{table}

\section{Summary}
\label{sec:conclusions}

In this work we have presented a determination of the parameters of resonances that 
appear in $K\pi$ scattering below 1.8 GeV. This has been achieved by means of 
series of Pad\'e approximants, 
which should converge to the appropriate analytic structure of the amplitude in a given domain. 
This constitutes another instance of the applicability and usefulness of this method, 
which avoids specific
model assumptions in the determinations of the mass, width, and coupling of a resonance. 
As a matter of fact, these parameters are usually obtained from Breit-Wigner-like parameterizations (or slight modifications) which
make a specific relation between the width and residue, and usually assume that the data contain
simple backgrounds superimposed
to the resonance signal.  With this method we determine the pole without such assumptions.
Moreover, it should be remarked that this method can be applied in the inelastic region, where the
powerful partial-wave dispersion relations cannot be used in practice to obtained poles.

In addition, these determinations have been obtained using as input a 
recent dispersive description of all 
the $K\pi$ data, which is constrained to satisfy two Forward Dispersion Relations 
(and several crossing sum rules) up to 1600 MeV. It should also be noted that simple 
fits to the data, as those used in previous determinations of resonance parameters, 
do not fulfill these fundamental constraints. These constrained fits
have also taken into account systematic uncertainties due to incompatibilities 
between different experiments.

Thus, we have provided
determinations of the mass, width and coupling to $K\pi$ for 
the conflictive $K_0^*(800)$ or $\kappa$ resonance,
the $K_0^*(1430)$ scalar, the $K^*(892)$ and $K_1^*(1410)$ vectors,  the spin-two
$K_2^*(1430)$ as well as the spin-three $K^*_3(1780)$. The results are fairly 
competitive with the results on the Review of Particle Properties, although it 
should be noted that these results contain some estimation of systematic and 
theoretical uncertainties usually lacking in the literature.

\section*{Acknowledgments}
JRP and AR are supported by the Spanish Project FPA2014-53375-C2-2
and the Spanish Excellence network HADRONet FIS2014-57026-REDT.
The work of JRE was supported by
the DFG (SFB/TR 16, ``Subnuclear Structure of Matter'')
and the Swiss National Science Foundation. AR would also like to acknowledge the ﬁnancial support of the Universidad Complutense de Madrid through a predoctoral scholarship.

\end{document}